\documentclass[epsf]{emulateapj}
\usepackage{natbib}

\newcommand{\ha}{H$\alpha$}

\newcommand{\Msun}{M$_{\sun}$}

\newcommand{\um}{$\mu$m}
\newcommand{\hii}{\ion{H}{2}}

\newcommand{\pa}{P$\alpha$}
\newcommand{\Lsun}{L$_{\sun}$}

\renewcommand{\um}{$\mu$m}
\renewcommand{\hii}{\mbox{\ion{H}{2}}}

\newcommand{\peryr}{yr$^{-1}$}

\begin{document}

\title{Toward a Unification of Star Formation Rate Determinations in the Milky Way and Other Galaxies}

\author{Laura Chomiuk\altaffilmark{1}$^{,}$\altaffilmark{2} \& Matthew S. Povich\altaffilmark{3}}
\altaffiltext{1}{National Radio Astronomy Observatory, P.O. Box O, Socorro, NM 87801; lchomiuk@cfa.harvard.edu}
\altaffiltext{2}{Harvard--Smithsonian Center for Astrophysics, MS-66, 60 Garden Street, Cambridge, MA 02138}
\altaffiltext{3}{NSF Astronomy and Astrophysics Postdoctoral Fellow in the Department of Astronomy and Astrophysics, The Pennsylvania State University, 525 Davey Lab, University Park, PA 16802; povich@astro.psu.edu}

\begin{abstract}
The star formation rate (SFR) of the Milky Way remains poorly known, with often-quoted values ranging from 1 to 10~M$_{\odot}$ yr$^{-1}$. This situation persists despite the potential for the Milky Way to serve as the ultimate SFR calibrator for external galaxies. 
We show that various estimates for the Galactic SFR are consistent with one another once they have been normalized to the same initial mass function (IMF) and massive star models, converging to $1.9\pm0.4$ M$_{\odot}$ yr$^{-1}$. 
However, standard SFR diagnostics are vulnerable to systematics founded in the use of indirect observational tracers sensitive only to high-mass stars. We find that absolute SFRs measured using resolved low/intermediate-mass stellar populations in Galactic \hii~regions are systematically higher by factors of ${\sim}2-3$ as compared with calibrations for SFRs measured from mid-IR and radio emission. We discuss some potential explanations for this discrepancy and conclude that it could be allayed if (1) the power-law slope of the IMF for intermediate-mass (1.5 M$_{\odot} < m < 5~$M$_{\odot}$) stars were steeper than the Salpeter slope, or 2) a correction factor was applied to the extragalactic 24~\um\ SFR calibrations to account for the duration of star formation in individual mid-IR--bright \hii\ regions relative to the lifetimes of O stars.
Finally, we present some approaches for testing if a Galactic SFR of ${\sim}2$ M$_{\odot}$ yr$^{-1}$ is consistent with what we would measure if we could view the Milky Way as external observers. Using luminous radio supernova remnants and X-ray point sources, we find that the Milky Way deviates from expectations at the 1--$3 \sigma$ level, hinting that perhaps the Galactic SFR is overestimated or extragalactic SFRs need to be revised upwards.
\end{abstract}


\section{Introduction}

Measuring the star formation rate (SFR) of a galaxy is like taking its pulse. The SFR quantifies the population of massive stars, which sets the composition and energetics of the interstellar medium (ISM) as well as the rate at which the ISM is dispersed through galactic winds and locked away into long-lived, low-mass stars. Our understanding of galaxy activity and the growth of stellar mass throughout cosmic time depends critically on measurements of SFR \citep[e.g.,][]{Madau_etal96, Bouwens_etal11}. Galaxy evolution models require the SFR as a fundamental input parameter \citep{Larson74, Chiappini_etal97, Kauffmann_Haehnelt00, deRossi_etal09}, since the SFR sets both the level of massive star feedback and the rate at which gas and dust are depleted.

The Milky Way is generally excluded from comparative studies of star-forming galaxies because we lack a unified, global picture of its structure and star formation activity. The location of the Solar System inside the Galactic disk gives us a disadvantaged perspective when attempting to study the Milky Way as a whole galaxy. Sightlines through the disk suffer very high extinction due to the dusty ISM, so SFR diagnostics which depend upon optical/UV observational tracers (e.g., \ha) cannot be applied to the Milky Way, just as they cannot be applied to external galaxies viewed nearly edge-on (see e.g., \citealt{Kennicutt98a} and \citealt{Calzetti_etal09} for reviews of SFR indicators). Radio and far-infrared (far-IR) observations are unaffected by extinction and therefore have provided most of the Galactic SFR estimates \citep[e.g.][]{Miller_Scalo79, Guesten_Mezger82, Misiriotis_etal06, Murray_Rahman10}. These wavelengths show great promise for providing convergence between extragalactic and Galactic SFR measurements, because a calibration like the \citet[][hereafter C+07]{Calzetti_etal07} 24~\um\ diagnostic for individual star-forming regions could be extended to the Galactic case. There remain, however, significant challenges stemming from uncertain distances to Galactic \hii\ regions and confusion of multiple star-formation regions overlapping along a given line of sight.

In spite of the observational challenges created by extinction, uncertain distances, and confusion, the Milky Way {\it should} be the ultimate SFR calibrator. The Galaxy offers the overwhelming advantage of high spatial resolution. Infrared observatories have advanced to the point that it is now routine to resolve individual stars in young embedded clusters several kpc distant. This makes it possible to directly probe the young stellar populations in massive clusters ionizing Galactic \hii\ regions, constraining the stellar initial mass function (IMF) and stellar ages and investigating the detailed physical processes governing star formation. The resulting detailed star formation histories of Galactic \hii\ regions and the associated molecular clouds can then be calibrated against diffuse emission tracers, like integrated IR luminosity, that are used to study star formation in external galaxies.

This paper seeks to answer three questions. (1) What are current estimates of the SFR of the Milky Way, and are they consistent with one another? Estimates have been presented in the literature for the past 35 years, ranging from ${\sim}1$~\Msun~\peryr\ \citep{Robitaille_Whitney10a} to ${\sim}10$~\Msun~\peryr\ \citep{Guesten_Mezger82}. Some of this variance is due to fundamental differences in the data, but, as we will demonstrate, systematics like the use of different IMFs or stellar models play a significant role. (2) How solidly are \emph{absolute} SFRs known? To relate SFRs to other physical quantities like gas masses and stellar masses, it is not enough to calibrate SFRs in a relative sense; we need to know that a measured SFR of 1 M$_{\odot}$ yr$^{-1}$ actually corresponds to 1 M$_{\odot}$ yr$^{-1}$ of material incorporated into stars. And finally, (3) Do Galactic SFR measurements give results consistent with extragalactic SFR diagnostics? Studies of the Galactic SFR generally use fundamentally different strategies, as compared with estimates of extragalactic SFRs; for example, H$\alpha$ is not a useful diagnostic in the Galactic plane because of high extinction, and most external galaxies are too distant to study their resolved stellar populations. We would like to be certain that, if we measure 1 M$_{\odot}$ yr$^{-1}$ of star formation in an external galaxy, the same conditions in the Milky Way would also produce a measured SFR of 1 M$_{\odot}$ yr$^{-1}$. We identify two SFR diagnostics which can be applied in both the Milky Way and external galaxies: a galaxy's most luminous supernova remnant (SNR) and its population of X-ray point sources.

Note that throughout this work we use the standard term ``Galactic SFR'' to refer to the SFR of the Milky Way, and ``extragalactic SFRs'' when referring to external galaxies. The subsequent sections of this paper are organized as follows: in \S\ref{inputs} we define the IMF and stellar population synthesis models used as input assumptions in our SFR determinations, and in \S\ref{mwdiag} we correct for these systematics in an effort to homogenize Galactic SFR estimates as much as possible. In \S\ref{sfrabs} we derive the SFR in a sample of Galactic \hii~regions from their resolved stellar populations and discuss the results in the context of systematic errors inherent in absolute SFR determinations. In \S\ref{extest} we present SFR tracers that can be self-consistently applied both to the Milky Way and to external galaxies. We summarize our conclusions in \S\ref{concl}.

\section{Our Assumed Inputs for Calculating SFRs\label{inputs}}
As all SFR diagnostics only probe a subset of stellar masses, the IMF is a critical assumption in SFR estimates. The IMF is often expressed as $\phi(\log{m}) = dN/d\log{m}$, the number of stars in the logarithmic mass interval between $\log{m}$ and $\log{m}+d\log{m}$. We normalize our results to the commonly-used ``broken power-law'' Kroupa IMF (\citealt{Kroupa_Weidner03}; used in the SFR calibrations of \citealt{Kennicutt_etal09}, hereafter K+09):
\begin{equation}
\phi(\log{m}) \propto m^{-\Gamma},
\end{equation}
\begin{equation}
\indent \Gamma = 0.3; \indent 0.1\ M_{\sun} < m < 0.5\ M_{\sun},
\end{equation}
\begin{equation} \label{eq:hi-mIMF}
\indent \Gamma = 1.3; \indent 0.5\ M_{\sun} < m < 100\ M_{\sun}.
\end{equation}
It is worth noting that the \citet{Chabrier05} IMF yields total stellar masses that are practically identical to the Kroupa IMF when both functions are normalized to high stellar masses (see Figure~\ref{IMF_cartoon}), implying that the SFRs calculated in this work should also represent SFRs assuming a Chabrier IMF. 

Most SFR diagnostics indirectly measure a specific electromagnetic signature of massive stars, usually Lyman continuum (hydrogen-ionizing) photon flux, and then extrapolate this to a larger stellar population. Therefore, to accurately compare SFRs, we must use the same stellar population models and Lyman continuum photon rates in all cases. Here, we will use the stellar population models employed in version 5.1 of Starburst99\footnote{http://www.stsci.edu/science/starburst99} \citep{Leitherer_etal99, Vazquez_Leitherer05} and used in the SFR calibration of K+09, assuming solar metallicity. These make use of the recent O star models of \cite{Smith_etal02} and \cite{Martins_etal05}. 

According to  \cite{Kennicutt_etal94} and \citet{Kennicutt98a}, a SFR of 1 M$_{\odot}$ yr$^{-1}$ produces a Lyman continuum photon rate $N_{c} = 9.26 \times 10^{52}$ photon s$^{-1}$ for the \citet{Salpeter55} IMF (assuming a mass range of 0.1--100 M$_{\sun}$). The ratio of SFR to Lyman continuum photon rate is 1.44 times smaller for the Starburst99/Kroupa IMF formulation than for the Salpeter IMF (K+09), so we will adopt the following relationship between Lyman continuum rate and SFR for the Starburst99/Kroupa IMF calibration: 
\begin{equation} \label{eq:nlyc}
\textrm{SFR}~[{\rm M_{\sun}~yr^{-1}}] = 7.5 \times 10^{-54}~N_{c}~[{\rm phot~s^{-1}}].
\end{equation}
This relationship will be used to compare Galactic SFR determinations throughout this paper.  

Equation~\ref{eq:nlyc} should be interpreted as the continuous SFR required to maintain a steady-state population of ionizing stars that produces an observed ionizing photon rate $N_c$. The ``steady state'' condition requires that the ionizing-star birth rate balance the death rate. The timescale tacitly assumed by Equation~\ref{eq:nlyc} to compute SFRs is thus the lifetime of the ionizing stars. Simple experiments with the Starburst99 code, assuming a constant SFR, demonstrate that on-going star formation must last a minimum time of ${\sim}8$~Myr (the lifetime of a late O star) to reach the steady-state value of $N_c$, and $N_c$ is within 5\% of the steady-state value by ${\sim}5$~Myr, the lifetime of an early O star \citep{Bertelli_etal94,Martins_etal05}. Unsurprisingly, O stars completely dominate the ionizing photon flux from actively star-forming populations. We caution that the steady-state approximation {\it underestimates} SFRs for stellar populations with durations of star formation significantly less than O-star lifetimes. We will bear this issue in mind throughout \S\ref{lycsfrs} and return to it in \S\ref{ages} below.

Where we determine SFRs from SN rates instead of Lyman continuum photon rates, we assume Padova isochrones \citep{Bertelli_etal94} and that stars in the mass range from 10--100~\Msun\ explode as SNe. To determine SFRs in individual star-forming regions from intermediate-mass star counts, we also assume a Kroupa IMF and use stellar population ages from the literature, which generally are based on pre-main-sequence (PMS) isochrones \citep{Siess_etal00}.

\section{Review of Milky Way Star Formation Rate Estimates} \label{mwdiag}
\subsection{Lyman Continuum Photon Rates}\label{lycsfrs}

\citet{Smith_etal78} established the ``canonical'' value for the Galactic SFR by calculating the total Lyman continuum photon rate required to maintain the ionization of all Galactic giant \hii~regions then known from radio continuum surveys. They reported $N_c = 4.7 \times 10^{52}~{\rm phot~s^{-1}}$ (shown by later studies to be a lower limit) based on the thermal free-free radio continuum flux emitted by the Galactic \hii\ regions in their sample, using the formulation in \citet{Mezger_etal74} and correcting for a factor of ${\sim}2$ internal absorption of Lyman continuum photons by dust. \citet{Smith_etal78} then used a \citet{Salpeter55} IMF (with limits of 0.1 and 100~\Msun), the massive star models of \citet{Panagia73}, and an (unrealistically low) assumed average duration of star formation in a Galactic radio \hii\ region of $5\times 10^5$~yr to transform from Lyman continuum photon rate to a Galactic SFR of ${\sim}5$~M$_{\sun}$~yr$^{-1}$ (factor of ${\sim}3$ reported uncertainty). Their measurement of $N_c$  translates to a Galactic SFR of 0.35~M$_{\sun}$~yr$^{-1}$ with the Starburst99/Kroupa IMF calibration used here (Table~\ref{tab:mwsfr}). This strikingly low value can largely be explained by the incompleteness of their \hii\ region sample, but is also contributed to by the inappropriateness of the Starburst99 steady-state calibration (Equation~\ref{eq:nlyc}) as applied to the young \hii\ regions in the Smith et~al.~sample (this will be discussed further in \S\ref{sfrdiff}).

Building upon the work of \citet{Smith_etal78}, \citet{Guesten_Mezger82} made a new estimate of the Galactic SFR from radio \hii\ regions, taking into account the thermal radio emission from ``extended low-density'' \hii\ regions and including Lyman continuum photons escaping from giant \hii\ regions. They measured $N_c = (2.7 \pm 0.8) \times 10^{53}~{\rm phot~s^{-1}}$, nearly six times the Smith et~al.~value, and reported a Galactic SFR of 11~\Msun~\peryr, which is clearly \emph{not} six times the SFR estimate of 5~\Msun~\peryr\ made by \citet{Smith_etal78}. This discrepancy is explained by the adoption of the lognormal IMF of \citet{Miller_Scalo79}, different limits on the IMF mass range (from 0.1 to 60~M$_{\sun}$), and different stellar models (see Appendix A of \citealp{Guesten_Mezger82} for a description of their conversion from Lyman continuum photon rate to stellar mass). In our homogenized Starburst99/Kroupa IMF calibration, the \citet{Guesten_Mezger82} measurement of $N_c$ translates to a SFR of $2.0\pm0.6$ M$_{\sun}$~yr$^{-1}$ (Table~\ref{tab:mwsfr}). The incorporation of low-density \hii\ regions means that this measurement included the contribution to ionization provided by older Galactic O stars omitted by \citet{Smith_etal78}, hence the steady-state calibration should be appropriate here.

\citet{Mezger87} calculated a slightly lower value for the Lyman continuum photon rate of $N_c = (2.1 \pm 0.6) \times 10^{53}~{\rm phot~s^{-1}}$, discrepant from their earlier value due to their use of R$_{\odot}$ = 8.5 kpc for the distance to the Galactic center (as opposed to R$_{\odot}$ = 10 kpc used by \citealt{Guesten_Mezger82}).
This discrepancy underscores the point that Galactic SFR measurements are often dependent on models of Milky Way structure and star formation therein. Here, we are not accounting for differences in the assumed Milky Way model. 
They also used a modified Miller--Scalo IMF with a less steep slope for the highest mass stars, and find a SFR of $5.7\pm1.6$ M$_{\sun}$~yr$^{-1}$. Using our Starburst99/Kroupa IMF calibration, we find a corresponding SFR of $1.6\pm0.5$ M$_{\sun}$~yr$^{-1}$.

\cite{Bennett_etal94} used {\it Cosmic Background Explorer (COBE)} observations of \ion{N}{2} 205~$\mu$m emission to measure the rate of Lyman continuum photons in the Milky Way, and correcting for dust absorption and photon escape, they found $N_c = (3.5 \pm 1.8) \times 10^{53}~{\rm phot~s^{-1}}$. Subsequently, \cite{McKee_Williams97} used the same {\it COBE} data and slightly different assumptions to measure a Lyman continuum photon rate of $N_c = (2.6 \pm 1.3) \times 10^{53}~{\rm phot~s^{-1}}$. These two estimates of N$_{c}$ agree with one another to within the 50\% uncertainties quoted by the authors, and are also in excellent agreement with the radio data from \citet{Guesten_Mezger82}.  

In a recent update to the original Galactic SFR estimates, \cite{Murray_Rahman10} used {\it Wilkinson Microwave Anisotropy Probe (WMAP)} observations of free-free emission to measure a dust-corrected Lyman continuum photon rate of $N_c = 3.2 \times 10^{53}~{\rm phot~s^{-1}}$. This is in good agreement with previous measurements of $N_{c}$ and corresponds to a SFR of 2.4 M$_{\odot}$ yr$^{-1}$ (Table~\ref{tab:mwsfr}). The {\it COBE} and {\it WMAP} studies utilize the integrated light of the Milky Way and average over its entire stellar population; therefore, the steady-state Starburst99 calibration is also appropriate to these studies. 

\subsection{Massive Star Counts and Supernova Rates}

Measurements of the core-collapse SN rate in the Milky Way have also been used to constrain the SFR. A SN rate of 0.85 events per century is expected for a SFR of 1 M$_{\odot}$ yr$^{-1}$, assuming a Kroupa IMF and a SN progenitor mass range of 10--100~M$_{\odot}$ (calculated using Starburst99 models with the parameters discussed in \S2). 

\cite{Reed05} surveyed the B2--O3 type stars (${\geq}$10~M$_{\odot}$) in the local Solar neighborhood (within 1.5 kpc) and assumed a model of the Milky Way to extrapolate these massive star counts to the entire Galaxy. He estimated a steady-state SN rate of 1--2 events per century, averaged over the lifetimes of massive stars ($\sim$20 Myr for 10~M$_{\odot}$). In our Starburst99/Kroupa IMF calibration, this translates to a SFR of 1.2--2.4 M$_{\odot}$ yr$^{-1}$. 

\cite{Diehl_etal06} used gamma-ray measurements of radioactive $^{26}$Al to constrain the core-collapse SN rate in the Milky Way to $1.9\pm1.1$ events century$^{-1}$, averaged over a timescale set by the half life of $^{26}$Al ($\sim 7.2 \times 10^{5}$ yr). They assumed a \citet{Scalo98} IMF with $\Gamma =1.7$ and a SN progenitor mass range of 10--120 M$_{\odot}$. Assuming massive star yields of $^{26}$Al as modeled by \cite{Limongi_Chieffi06}, the yields for a Kroupa IMF are a factor of 1.12 higher than for the Scalo IMF. This means that for our normalized assumptions, the SN rate from radioactive $^{26}$Al is $1.7\pm1.0$ events century$^{-1}$, consistent with the results from \cite{Reed05}. This translates to a SFR of $2\pm1.2$ M$_{\odot}$ yr$^{-1}$.

\begin{deluxetable*}{lccc}
\tabletypesize{\normalsize}
\tablewidth{0 pt}
\tablecaption{ \label{tab:mwsfr}
 Normalized SFR Estimates for the Milky Way}
\tablehead{Method & Observation & SFR & Reference \\
 & & (M$_{\odot}$ yr$^{-1}$) & }
\startdata
Ionization Rate & Radio Free-Free & 0.35 & \cite{Smith_etal78} \\
Ionization Rate & Radio Free-Free & $2.0\pm0.6$ & \cite{Guesten_Mezger82} \\
Ionization Rate & Radio Free-Free & $1.6\pm0.5$ & \cite{Mezger87} \\
Ionization Rate & COBE \ion{N}{2} 205 $\mu$m & $2.6\pm1.3$ & \cite{Bennett_etal94} \\
Ionization Rate & COBE \ion{N}{2} 205 $\mu$m & $2.0\pm1.0$ & \cite{McKee_Williams97} \\
SN Rate & O/B Star Counts & $1.8\pm0.6$ & \cite{Reed05} \\
SN Rate & INTEGRAL Gamma-ray $^{26}$Al & $2.0\pm1.2$ & \cite{Diehl_etal06} \\
Dust Heating & COBE FIR continuum & $1.9\pm0.8$\tablenotemark{a} & \cite{Misiriotis_etal06} \\
Ionization Rate & WMAP Free-Free & $2.4\pm1.2$\tablenotemark{a} & \cite{Murray_Rahman10} \\
YSO Star Counts & \emph{Spitzer} IR Photometry & $1.1\pm0.4$ & \cite{Robitaille_Whitney10a} \\
YSO Star Counts & \emph{MSX} IR Photometry & $1.8\pm0.3$ & \cite{Davies_etal11} \\
\enddata
\tablenotetext{a}{Assumed 50\% uncertainty, as no error is quoted by the original author.}
\end{deluxetable*}

\subsection{Infrared Diffuse Emission and Point Sources}

\citet{Misiriotis_etal06} constructed a model of the spatial distribution of stars, gas, and dust in the Galaxy constrained by {\it COBE} observations of the Galactic IR emission. They measured the 100 $\mu$m luminosity of the Milky Way and, using the conversion described in \cite{Misiriotis_etal04}, calculated a SFR of 2.7 M$_{\odot}$ yr$^{-1}$. However, they used a Salpeter IMF (0.1--100 M$_{\odot}$) and \cite{Fioc_Rocca97} stellar population models. We attempt to correct for the difference in IMF by dividing the SFR by 1.44 (as above, see Equation \ref{eq:nlyc}), assuming that the Lyman continuum photons and the non-ionizing UV photons which produce FIR emission scale together (a good assumption, as the disparity between the Salpeter and Kroupa IMFs occurs at low stellar masses). Typically, galaxy-wide mid-IR measurements of SFR are sensitive to star formation over longer timescales than other SFR diagnostics ($\sim$100 Myr; C+07), as even non-ionizing B stars can contribute significantly to dust heating \citep{Cox_etal86}. We note that FIR diagnostics of the SFR include a host of critical assumptions, like what fraction of stellar light is absorbed by dust and how much older stellar populations contribute to dust heating. Still, the estimate of Galactic SFR from \cite{Misiriotis_etal06} is consistent with the others described here, at 1.9 M$_{\odot}$ yr$^{-1}$ (Table~\ref{tab:mwsfr}). 

\cite{Robitaille_Whitney10a} produced Galactic population synthesis models of young stellar objects (YSOs) that still retain dusty circumstellar disks and/or natal envelopes, assuming Kroupa IMF, a simple prescription for accretion, and a YSO lifetime of 2 Myr. These population synthesis models were then used to simulate the \citet{Robitaille_etal08} catalog of YSOs identified via mid-IR excess emission from the \emph{Spitzer Space Telescope} GLIMPSE (Galactic Legacy Infrared Mid-Plane Survey Extraordinaire; \citealt{Churchwell_etal09}) surveys of the Galactic plane. From comparing the simulations to the catalog, \citet{Robitaille_Whitney10a} derived a SFR of 0.68--1.45 M$_{\odot}$ yr$^{-1}$. 
\cite{Robitaille_Whitney10a} note that their quoted range of SFRs includes only the uncertainties on the total number of YSOs in the \citet{Robitaille_etal08} catalog.

\cite{Davies_etal11} modeled massive ($m>8$~\Msun) YSOs found in the Red \emph{MSX} Sources (RMS) Survey, using the data to find the best fits to basic YSO parameters such as bolometric luminosity and mass. This study assumed a massive YSO lifetime of $\sim$0.1 Myr and a Kroupa IMF, resulting in 
a Galactic SFR of 1.5--2.0 M$_{\odot}$ yr$^{-1}$.

\subsection{Galactic SFR Determinations Compared}

We see remarkably good agreement in Table~\ref{tab:mwsfr} between estimates of the Galactic SFR (with the exception of the oldest, \citealt{Smith_etal78}, which was clearly incomplete and superseded by \citealt{Guesten_Mezger82}), after we have taken care to normalize assumptions about the parent stellar populations. The estimates cluster around an average of $1.9\pm0.4$ M$_{\odot}$ yr$^{-1}$ (excluding \citealt{Smith_etal78} and \citealt{Guesten_Mezger82}, which were both improved upon by \citealt{Mezger87}). The most significant outlier is the result from \cite{Robitaille_Whitney10a}. We note that their Galactic SFR could be regarded as a lower limit, due both to issues of completeness and to the 2~Myr lifetime of the YSO phase used in their preliminary population synthesis models. This age limit is likely an overestimate, because it is based on circumstellar disk lifetimes for low-mass PMS stars, while the \citet{Robitaille_etal08} catalog is sensitive primarily to intermediate-mass YSOs (masses ${>}3$~\Msun). There is mounting evidence that the optically thick inner disks of intermediate-mass YSOs are rapidly destroyed over timescales ${<}1$~Myr \citep{Povich_Whitney10,Koenig_Allen11,Povich_etal11}. Such a correction in a future update of the \citet{Robitaille_Whitney10a} population synthesis models could easily provide the factor of ${\sim}2$ increase necessary to bring the Galactic SFR from intermediate-mass YSO star counts into agreement with the other estimates.

We stress that our homogenized estimate is significantly different from the Galactic SFR of 4--5~\Msun~yr$^{-1}$ often quoted in the literature, and this  disparity is due only to differences in the IMF and stellar models used. Investigators hoping to compare properties of the Milky Way with other galaxies \citep[e.g., the luminosity functions of star clusters;][]{Hanson_Popescu08} must first calibrate the SFRs to the same scale. The Starburst99/Kroupa IMF calibration is likely to be used often in the future, in which case the corresponding Galactic SFR is 2~M$_{\odot}$~yr$^{-1}$.

\section{Calibrating SFR Diagnostics Against Direct Star Counts}\label{sfrabs}

We have demonstrated that various measurements of the Galactic SFR using myriad observational tracers converge to a consistent value, provided that systematic differences are minimized by adopting a uniform IMF and stellar models.
We have made no assertions that our assumed inputs for calculating SFRs (\S\ref{inputs}) are the correct ones; they were chosen simply to match the Starburst99 stellar population synthesis models widely applied to SFR determinations in external galaxies. If these assumptions were incorrect, then {\it relative} SFR comparisons would be unaffected, but potentially large systematic errors would be introduced into {\it absolute} SFR determinations.

\begin{figure}[t]
\epsscale{1.2}
\plotone{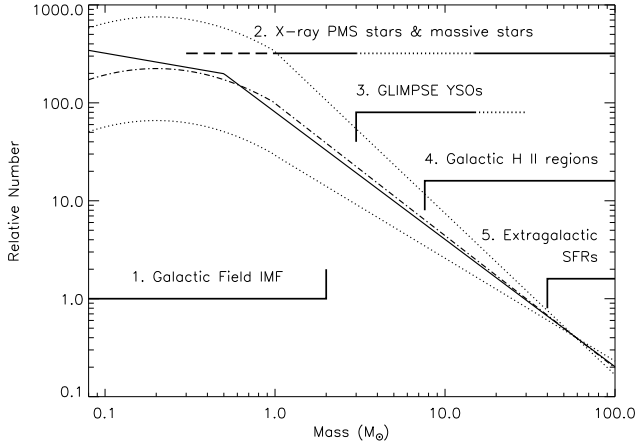}
\caption{Simulated Galactic-field IMF \citep[solid curve;][]{Kroupa_Weidner03} and lognormal+power-law IMF \citep[dash-dotted curve;][]{Chabrier05} normalized to the same relative numbers of early O-type stars ($m>40$~\Msun). 
Dotted curves show an error envelope on the Chabrier IMF based on an uncertainty of 0.3 in the value of the power-law slope for $m>1$~\Msun.
Five mass ranges (numbered) relevant to multi-wavelength determinations of the IMF and SFRs are indicated schematically by brackets and discussed in the text. 
\label{IMF_cartoon}}
\end{figure}

While the general broken-power-law (or lognormal+power-law; \citealp{Chabrier03}) form of the observed IMF is remarkably consistent across stellar populations in widely disparate environments, variations have been reported both in the power-law slopes and in the locations of the break points where the slopes change \citep{Kroupa01,Bastian_etal10}.  Although these variations have low significance given the typically large uncertainties inherent in IMF measurements \citep{Bastian_etal10}, even an insignificant deviation from the default Salpeter--Kroupa slope can have significant consequences.  The center-of-mass of the IMF, defined as the point where the integrated stellar mass is divided equally between higher and lower masses, falls near 1~\Msun\ \citep{Kroupa01}. If we estimate SFRs by extrapolating over an IMF that is pinned only to massive stars, we maximize the impact of IMF variations because such massive stars are farthest from the center-of-mass. This effect is illustrated in Figure~\ref{IMF_cartoon}, in which the lognormal+power-law ($\Gamma = 1.35$) IMF of \citet{Chabrier05} is plotted alongside the standard Kroupa IMF. Both IMF models have been normalized to the same relative numbers of very high-mass stars (early O stars with $m > 40$~\Msun, which dominate the bolometric luminosity and Lyman continuum photon flux in observable extragalactic star-forming regions; \citealt{Martins_etal05}). 
With this normalization, the Chabrier IMF yields 15\% fewer stars (defined henceforth as $N_{\rm pop}$) but 3\% greater integrated stellar mass ($M_{\rm pop}$) above the hydrogen-burning limit ($m\ge 0.08$~\Msun) compared to the Kroupa IMF. The insignificant difference in $M_{\rm pop}$ means that the choice between these IMFs is unimportant for deriving SFRs from unresolved stellar populations in \hii~regions. However, an uncertainty of 0.3 on the value of $\Gamma$ \citep[e.g.][]{Muench_etal02,Chabrier05} increases $N_{\rm pop}$ by a factor of 2.3 and $M_{\rm pop}$ by a factor of 1.6 on the high side envelope (upper dotted curve in Figure~\ref{IMF_cartoon}).

High-resolution, wide-field, multi-wavelength imaging observations enable us to derive SFRs in bright Galactic \hii\ regions (the nearby analogs of observable extragalactic star-forming regions) using {\it direct} detections of the resolved low- and intermediate-mass stellar populations. Such an approach samples the center of the IMF, greatly reducing the uncertainty of extrapolation (Figure~\ref{IMF_cartoon}) while avoiding altogether any dependence on models of massive star properties. Hence the following analysis provides an independent estimate of the systematic errors on absolute SFR calibrations.

\subsection{The Galactic \hii\ Region M17 as a Case Study} \label{m17}

\subsubsection{The X-ray Luminosity Function: SFR from Star Counts} \label{m17xlf}

We begin by considering the well-studied, nearby (2.0 kpc; \citealt{Xu_etal11}) Galactic \hii\ region M17. Because of its relatively high obscuration ($A_V>8$~mag), bright nebulosity, and overwhelming field-star contamination produced by its location in the inner Galactic plane ($[l,b]=[15\fdg1,-0\fdg7]$), establishing a reliable membership catalog for NGC 6618, the massive cluster ionizing M17, has proven prohibitively difficult using optical/IR photometry alone (see discussions in \citealp{Broos_etal07} and \citealp{Povich_etal09}). These problems have been overcome by employing high-resolution imaging spectroscopy from the Advanced CCD Imaging Spectrometer (ACIS) of the {\it Chandra X-ray Observatory} to identify cluster members \citep{Broos_etal07}. Low- and intermediate-mass PMS stars exhibit powerful convectively-driven magnetic reconnection flaring activity, hence they are $10^2$--$10^3$ times more luminous in X-rays than older Galactic field stars, and their X-ray luminosity $L_X$ is broadly correlated with bolometric luminosity and hence stellar mass \citep{Preibisch_Feigelson05}. Hard (${>}2$~keV) X-ray photons are relatively unaffected by interstellar absorption, and while optical/IR photometry in bright \hii~regions is plagued by diffuse nebular emission, ACIS point-source detections generally do not suffer significant reductions in sensitivity due to diffuse emission \citep{Broos_etal07}. 

The stellar X-ray luminosity function (XLF; the distribution of stars as a function of $L_X$) itself provides a good statistical tracer of the underlying IMF in young clusters \citep{Getman_etal06, Wang_etal07}. The approximate range of masses sampled by observed XLFs of various massive clusters is indicated in Figure \ref{IMF_cartoon}. In addition to PMS stars, X-ray observations are highly sensitive to massive stars, which produce strong X-ray emission via microshocks in their stellar winds and a variety of more exotic mechanisms \citep{Gagne_etal11}. Intermediate-mass A- and B-type stars on or near the zero-age main sequence lack both magnetic dynamos and powerful stellar winds, hence they are generally X-ray quiet \citep[][and references therein]{Povich_etal11}.

To estimate $N_{\rm pop}$ from the XLF of a given cluster, the simplest approach \citep[e.g.,][]{Feigelson_etal11} is to use as a calibrator the XLF of the ONC, which contains 839 lightly-obscured (X-ray median energy $<2$~keV), low-mass members \citep{Getman_etal05}. The XLF in the target cluster is compared to the calibrator ONC XLF to derive a scale factor, which is then multiplied by $N_{\rm pop}({\rm ONC})$, the population of the ONC above the brown-dwarf limit \citep{Hillenbrand_Hartmann98}.  

Estimates of $N_{\rm pop}$ from XLF scaling assume several fundamental similarities between the target cluster and the ONC: (1) the form of the underlying IMF, (2) the ratio of obscured/unobscured cluster members, (3) the circumstellar disk fraction of the cluster membership ($L_X$ is lower for protostars and YSOs with disks compared to diskless PMS stars of the same mass; see \citealp{Povich_etal11} and references therein), and (4) the transformation between XLF and IMF. With the possible exception of (1), none of these assumptions is generally valid. Assumptions (3) and (4) could hold if the age of the target cluster is sufficiently close to that of the ONC, but in practice ages are difficult to determine, and in any case it is not yet known how much the target cluster could differ from the ONC age before producing a significant impact on the XLF scaling. To avoid assumption (2), a ``two-component'' approach can be employed, in which the cluster XLF is divided into lightly-obscured and heavily-obscured sub-populations, and the calibrator ONC XLF is scaled independently to the XLF of each sub-population \citep[this second approach was used by][]{Broos_etal07}.

In M17, the XLF was measured from 886 sources detected by {\it Chandra}/ACIS \citep{Broos_etal07}. \cite{Povich_etal09} showed that $N_{\rm pop}$ as reported by \citet{Broos_etal07} for NGC 6618 translates to a SFR based on star counts of SFR$_{\rm SC} = 0.008$--0.01~\Msun~\peryr, using the \citet{Muench_etal02} IMF to convert $N_{\rm pop}$ to $M_{\rm pop}$. Adopting instead our standard Kroupa IMF and an upper envelope of 1.5~Myr (see discussion in \citealt{Povich_etal09}) for the age spread of the X-ray-detected stellar population in NGC 6618, SFR$_{\rm SC}$ becomes 0.004--0.006~\Msun~\peryr. 

In the following two subsections, we estimate the M17 SFR using independent tracers of the massive stellar population and compare the results to SFR$_{\rm SC}\sim 0.005$~\Msun~\peryr.

\subsubsection{The Thermal Radio Continuum SFR Diagnostic in M17} \label{m17radio}

Historically there has been some disagreement between different measurements of the M17 radio continuum flux density. Adopting $S_{\rm 5Ghz}=784$~Jy and $S_{\rm 15Ghz}=609$~Jy from \cite{Povich_etal07} and using the same conversion from thermal continuum (free-free emission) to Lyman continuum photon rate as \citet{Smith_etal78}, we calculate $N'_c=3.07\times 10^{50}$~s$^{-1}$ at 5 GHz and $2.66\times 10^{50}$~s$^{-1}$ at 15~GHz. The difference reflects a ${\sim}15\%$ uncertainty on $N'_c$. 

Summing over the known O stars in NGC 6618 and using the models of \citet{Martins_etal05} to obtain the Lyman continuum photon emission rates
as a function of spectral type, the combined Lyman continuum photon rate emitted is $N_c = 3\times 10^{50}$~s$^{-1}$. This is in surprisingly good agreement with $N'_c$ derived above, given that some fraction of the Lyman continuum photons are expected to be absorbed by dust before contributing to the ionization of the \hii\ region, and some fraction of the Lyman continuum flux likely escapes from the nebula altogether \citep{Povich_etal07}. Our calculation of $N_c$ largely neglected the contribution of stars later than O8 to the ionization, but the inclusion of O8--B3 stars would only increase $N_c$ by a few percent. There may yet be new major ionizing stars waiting to be discovered in M17 (\citealp{Povich_etal08} reported the discovery of a new candidate O5 V star), but it is unlikely that a significant fraction of the ionizing stellar population in M17 has been missed. We have accounted for known massive binaries among the majority of the principal ionizing stars (types O6 and earlier) in our calculation of $N_c$ \citep{Povich_etal09}. We note that the total bolometric luminosity of the known O stars is $L_{\rm OB}\sim 5.8\times 10^6$~\Lsun, in very good agreement with the luminosity in the global spectral energy distribution (SED) of the M17 nebula \citep{Povich_etal07}.  

Absorption of Lyman continuum photons by dust has a negligible impact upon the ionization of M17. Somewhat more important is the escape of Lyman continuum photons into the diffuse ISM through the blow-out in the \hii\ region \citep{Povich_etal07}. In the most probable nebular geometries, $N'_c/N_c \ga 0.9$ \citep{Povich_etal07}. Given the uncertainty in $N'_c$, it is not unreasonable to assume $N_c=N'_c=3\times 10^{50}$~s$^{-1}$ for this highly embedded \hii~region. Using our normalized conversion in Equation \ref{eq:nlyc}, this translates to SFR$_{ff}=0.0022$~\Msun~\peryr\ traced by free-free emission, which is {\it lower} than SFR$_{\rm SC}$ by a factor of ${\ga}2$. In \S\ref{sfrdiff} we will discuss some possible explanations for this discrepancy. 

\subsubsection{The 24~\um\ Mid-IR SFR Diagnostic in M17} \label{m17ir}

According to C+07, their {\it Spitzer} 24~\um\ SFR calibration ``is appropriate for metal-rich \hii\ regions or starbursts'' where the radiation field is dominated by young massive stars. M17 is a highly embedded Galactic \hii\ region that radiates ${\ga}90\%$ of its luminosity in the mid-IR \citep{Povich_etal07}, hence it meets these criteria. We can therefore use M17 to test the applicability of the C+07 extragalactic mid-IR SFR diagnostic to a nearby, resolved \hii\ region.

M17 was observed using the Multiband Imaging Photometer (MIPS) on board {\it Spitzer} as part of the MIPSGAL survey \citep{Carey_etal09}, but because the \hii\ region is bright enough to completely saturate even the short MIPSGAL exposures at 24~\um, there is no direct measurement of the luminosity in the MIPS 24~\um\ band, $L_{24}$, that can be used as an input to the C+07 SFR relation (their Equation 9). Instead, we interpolate the global SED presented by \citet{Povich_etal07} over the range of 20 to 60~\um\ and compute the 24~\um\ flux density, $F_\nu(24)$ (see Figure~\ref{SFRcal}{\it a}). This gives
\begin{equation}\label{L24}
  L_{24} \equiv \nu L_\nu(24) = 4\pi d^2 \nu F_\nu(24) = 6.6\times 10^{39}~{\rm erg~s^{-1}},
\end{equation}
(${<}25\%$ uncertainty, dominated by the uncertain distance $d$) and, following C+07, SFR$_{24}=$SFR$_{ff}=0.0022$ \Msun~\peryr, 
just as we derived from radio free-free emission above. We therefore find that SFRs based on independent indirect tracers of Lyman continuum photon rate are in agreement, but both are lower than SFR$_{\rm SC}$ by factors of ${\ga}2$. We note that the C+07 calibration uses the Kroupa IMF and Starburst99 models as we have used throughout this paper, including in our calculation of SFR$_{ff}$. As cautioned in \S\ref{inputs}, M17 may be too young for the steady-state approximation to hold. We will discuss the consequences of ignoring this caveat in \S\ref{ages}.

\begin{deluxetable*}{llcccccccc}
\tabletypesize{\scriptsize}
\tablewidth{0 pt}
\tablecaption{ \label{G8}
SFR Data for 8 Galactic \hii~Regions}
\tablehead{
\colhead{} & \colhead{Alternate} & \colhead{d} & \colhead{Earliest} & \colhead{$M_{\rm pop}$} & \colhead{$\tau_{\rm SF}$} & \colhead{$\log{L_{24}}$} & \colhead{$\log{N^{\prime}_c}$} & \colhead{SFR$_{\rm SC}$} & \colhead{} \\
\colhead{Name} & \colhead{Name} & \colhead{(kpc)} & \colhead{known star(s)} & \colhead{(\Msun)} & \colhead{(Myr)\tablenotemark{a}} & \colhead{(erg s$^{-1}$)\tablenotemark{b}} & \colhead{(phot s$^{-1}$)\tablenotemark{c}} & \colhead{($10^{-3}$~\Msun~\peryr)} & \colhead{Refs.\tablenotemark{d}}
}
\startdata
Carina    & NGC 3372  & $2.3\pm  0.1$ & LBV,WNL,O2 I     & $43000\pm 5000$ & 5   & $40.04\pm 0.04$ &	$50.9\pm 0.1$ & 8.6      & 1,2 \\
RCW 49\tablenotemark{e} & Wd 2      & $4.2\pm 1.5$  & $4\times$WNL,O3 V & ${>}4000$	  & 3   & $39.98\pm 0.35$ & $50.5\pm 0.3$ & ${>}1.3$ & 3,4 \\
M17       & NGC 6618  & $2.0\pm 0.1$  & $4\times$O4 V    & $8000\pm 1000$  & 1.5 & $39.82\pm 0.08$ &	$50.5\pm 0.1$ & 5.3      & 5,6,7 \\
NGC 6357  & Pismis 24 & $1.7\pm 0.5$  & O3.5 I, O3.5 III     & $6500\pm 1300$  & 2   & $39.24\pm 0.20$ &	$49.6\pm 0.3$ & 3.2      & 8,9,10 \\
RCW 38 &           & $1.7\pm 0.2$  & O5.5 V & $1200\pm 300$   & 1   & $39.03\pm 0.10$ & $49.8\pm 0.1$ &	1.2      & 11 \\ 
W3 Main   &           & $2.0\pm0.07$  & Several $\sim$O5      & $3250\pm 1300$  & 1.5 & $38.93\pm 0.03$ &     $49.45\pm 0.05$ &  2.2      & 12 \\
Rosette   & NGC 2244  & $1.3\pm0.1$  & O4 V & $1300\pm 200$ & 2   & $38.63\pm 0.10$ &  $49.8\pm 0.1$   &	0.65     & 13,14 \\
ONC & M42       & $0.41\pm 0.04$& O7 V & $1600\pm 100$ & 3   & $38.48\pm 0.08$ &	$48.77\pm 0.09$ &	0.53     & 15,16 \\
\enddata
\tablenotetext{a}{The star formation timescale $\tau_{\rm SF}$ is defined as the age of the {\it oldest} stars included in the IMF
  scaling that yielded $M_{\rm pop}$. Hence SFR$_{\rm SC}$ is the
  time-averaged SFR over $\tau_{\rm SF}$, and the actual
  SFR during any short bursts could be significantly higher.}
\tablenotetext{b}{$L_{24}\equiv \lambda L_\lambda$ is the equivalent mid-IR continuum luminosity in the MIPS 24~\um\ bandpass.}
\tablenotetext{c}{$N_c^\prime$ is the Lyman continuum photon rate required to maintain the ionization of the radio H~II~region, following the notation and method of calculation employed by \citet{Smith_etal78}.}
\tablenotetext{d}{1 = \citet{Smith_Brooks07}, 2 = \citet{Povich_etal11}, 3 = \citet{Tsujimoto_etal07}, 4 = \citet{Ascenso_etal07}, 5 = \citet{Povich_etal07}, 6 = \citet{Broos_etal07}, 7 = \citet{Povich_etal09}, 8 = \citet{Wang_etal07}, 9 = \citet{Russeil_etal10}, 10 = \cite{Maiz_Apellaniz_etal07}, 11 = \citet{Wolk_etal06}, 12 = \citet{Feigelson_Townsley08}, 13 = \citet{Celnik85}, 14 = \citet{Wang_etal08}, 15 = \citet{Hillenbrand_Hartmann98}, 16 = \citet{Feigelson_etal02}.}
\tablenotetext{e}{$M_{\rm pop}$ and SFR$_{\rm SC}$ are lower limits for RCW 49 because the 2.8~kpc distance \citep{Ascenso_etal07} assumed for the IMF scaling is at the near end of the wide range of distance estimates reported in the literature.}
\end{deluxetable*}

The C+07 $L_{24}$ SFR diagnostic is a statistical relation based on 179 observed \hii\ region ``knots'' in 23 different high-metallicity galaxies, and its disagreement with SFR$_{\rm SC}$ in a single, Galactic \hii\ region is hardly a cause for alarm. Perhaps M17 is simply an outlier from the trend, or perhaps there is some fundamental difference in measurement between the Galactic and extragalactic observations. \citet{Povich_etal07} measured the mid-IR luminosity of M17 using a large aperture chosen to match the linear extent of the \hii\ region and photodissociation region, while the \hii\ knots in distant galaxies were generally smaller than the apertures used by C+07 to extract their fluxes. Hence, there could be an offset between the mid-IR luminosity measured for M17 and the luminosities of the extragalactic \hii\ knots due to the drastically different filling factors and background subtractions involved. The impact of such potential biases can be evaluated by hypothetically locating M17 among the C+07 extragalactic sample (their Figure~9). For each \hii\ knot in their sample, C+07 measured ``luminosity surface densities'' (LSDs; units of erg~s$^{-1}$~kpc$^{-2}$) $S_{\rm P\alpha}$ and $S_{24}$ in the P$\alpha$ recombination line and the MIPS 24~\um\ band, respectively,  using a fixed photometric aperture matched to the MIPS 24~\um\ resolution. The aperture size corresponded to a linear scale of ${\sim}300$~pc at $d=5$~Mpc. 

Among the high-metallicity \hii\ knots, the range of LSDs observed by C+07 was $38\la \log{S_{\rm P\alpha,corr}}\la 40$ and $40.5\la \log{S_{24}}\la 42.5$. An \hii\ region with the same $L_{24}$ as M17, located in a galaxy 5~Mpc distant, would have an observed LSD of $\log{S_{24}}=41.00$, well within the range. We can calculate the luminosity $L_{\rm P\alpha}$  in the \pa\ line from the thermal radio continuum (\S\ref{m17radio}), expressed as: 
\begin{equation}
  L_{\rm P\alpha}=4\pi d^2\frac{j({\rm P\alpha})}{j_\nu(ff)}F_\nu(\nu).
\end{equation}
where $j({\rm P\alpha})$ and ${j_\nu(ff)}$ are the P$\alpha$ and free-free emissivities \citep{Osterbrock74, Giles77} and $F_\nu(\nu)$ is the free-free radio continuum flux density at frequency $\nu$ (see \citealt{Povich_etal07} for more detail). This gives $L_{\rm  P\alpha}=4.1\times 10^{37}$~erg~s$^{-1}$. Assuming again $d=5$~Mpc and a 300~pc test aperture, $\log{S_{\rm  P\alpha}}=38.76$, also within the range of values reported by C+07 (since it is based on radio observations, there is no need to correct the LSD for extinction).  

Although M17 is a relatively low-luminosity \hii~region in the extragalactic context, there is nothing unusual about its LSDs compared to the C+07 sample. An M17 analog would be observable in the SINGS \citep[\emph{Spitzer} Infrared Nearby Galaxies Survey;][]{SINGS} 24~\um\ images as an \hii\ knot in a nearby galaxy, assuming that it were not confused with a neighboring, brighter \hii~region, and it would fall well within the LSD distributions measured by C+07 for high-metallicity \hii~regions.

\subsection{Comparison of Mid-IR Galactic and Extragalactic SFR Calibrations}\label{g-ex}

We now generalize the analysis of M17 described above to a sample of eight Galactic \hii~regions (Table~\ref{G8}). With one exception, this sample consists of \hii~regions that, like M17, meet all of the following criteria: (I) ionized by at least one early O-type star (spectral type O5 V or earlier); (II) sufficiently embedded 
that the bulk of the bolometric luminosity is reprocessed by dust and emitted in the IR; (III) ionizing cluster(s) with well-sampled (complete to $m\la 1.5$~\Msun) XLFs reported in the literature; and (IV) evidence for ongoing star formation (for example, a detected population of stars with IR-excess emission from disks). The exception is the ONC, which lacks early O-type stars but was included because it serves as an observational touchstone for massive cluster studies in general, and XLF studies in particular. Table~\ref{G8} contains data relevant to SFR determinations for each \hii~region: 

{\it Mass of stellar population.} The total number of stars $N_{\rm pop}$ associated with the ionizing cluster(s) was adopted from published XLFs (see references listed in Table~\ref{G8}) and converted to total stellar mass, $M_{\rm pop}$, for $m\ge 0.1$~\Msun\ using the Kroupa IMF. Uncertainties on $M_{\rm pop}$ reflect only the statistical uncertainties on $N_{\rm pop}$, which include both the uncertain distance to each cluster and uncertainties introduced in extrapolating the XLFs to correct for incompleteness. Because we cannot always be certain that the XLF of a given cluster is complete, it may be safest to regard our reported $M_{\rm pop}$ values as lower limits. 

{\it Star formation timescale and SFR$_{SC}$.} Following our selection criterion (IV) above, we assume that star formation has been continuous over the lifetime of each \hii\ region, defined as the star formation timescale $\tau_{\rm SF}$.
While X-ray emission gives no useful constraint on the ages of the stars, follow-up optical and IR photometry of the X-ray detected stellar population does. Values of $\tau_{\rm SF}$ for each region are based on published results from isochrone fitting to optical/near-IR color-magnitude diagrams (CMDs) and/or estimates of circumstellar disk fractions using observed IR excess emission.  Age determinations for very young (${\sim}1$~Myr) stellar populations in regions with ongoing star formation are notoriously imprecise, since the observed age spreads are generally comparable to the measurement uncertainties and the reported ages themselves. We therefore use the upper end of the range of ages reported in the recent literature (see references listed in Table~\ref{G8}) for each cluster. We had three motivations for this choice: (1) The difficulty in correcting for differential reddening in near-IR CMDs makes it easier to determine an upper age envelope than an average age for an obscured cluster. (2) Our values of $M_{\rm pop}$ may only be lower limits in some cases. (3) Our analysis of M17 (\S\ref{m17}) indicates that, in spite of a ${\sim}50\%$ uncertainty on the age, SFR$_{\rm SC}$ is significantly higher than SFR based on diffuse tracers, hence we need only to place {\it lower} limits on the time-averaged SFR$_{\rm SC}=M_{\rm pop}/\tau_{\rm SF}$ (Table~\ref{G8}) to test the generality of this result. We note that if there are bursts of higher SFR during the lifetime of an \hii\ region, this would (temporarily) increase SFR$_{\rm SC}$.

\begin{figure*}[t]
\epsscale{1.1} 
\plottwo{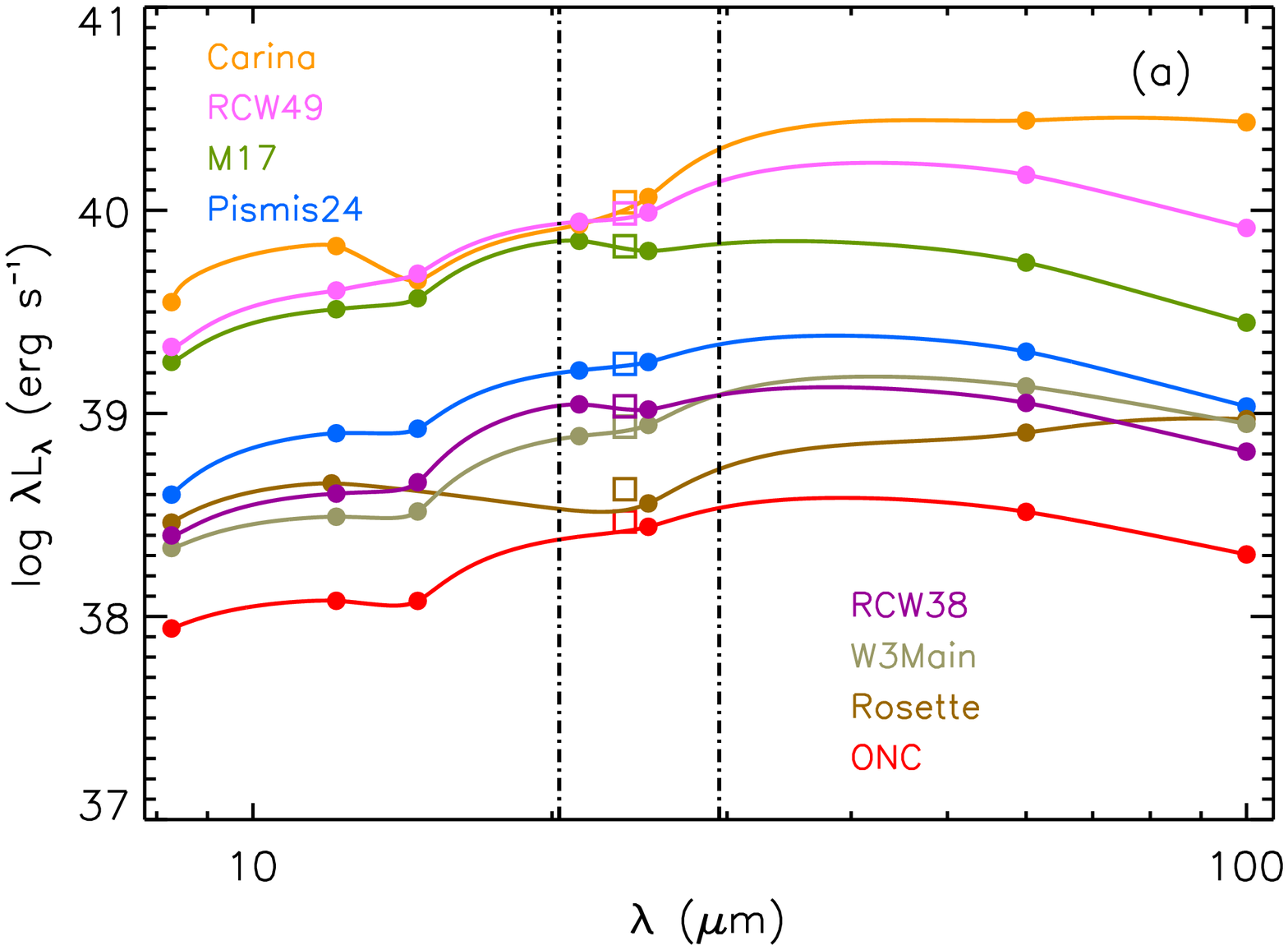}{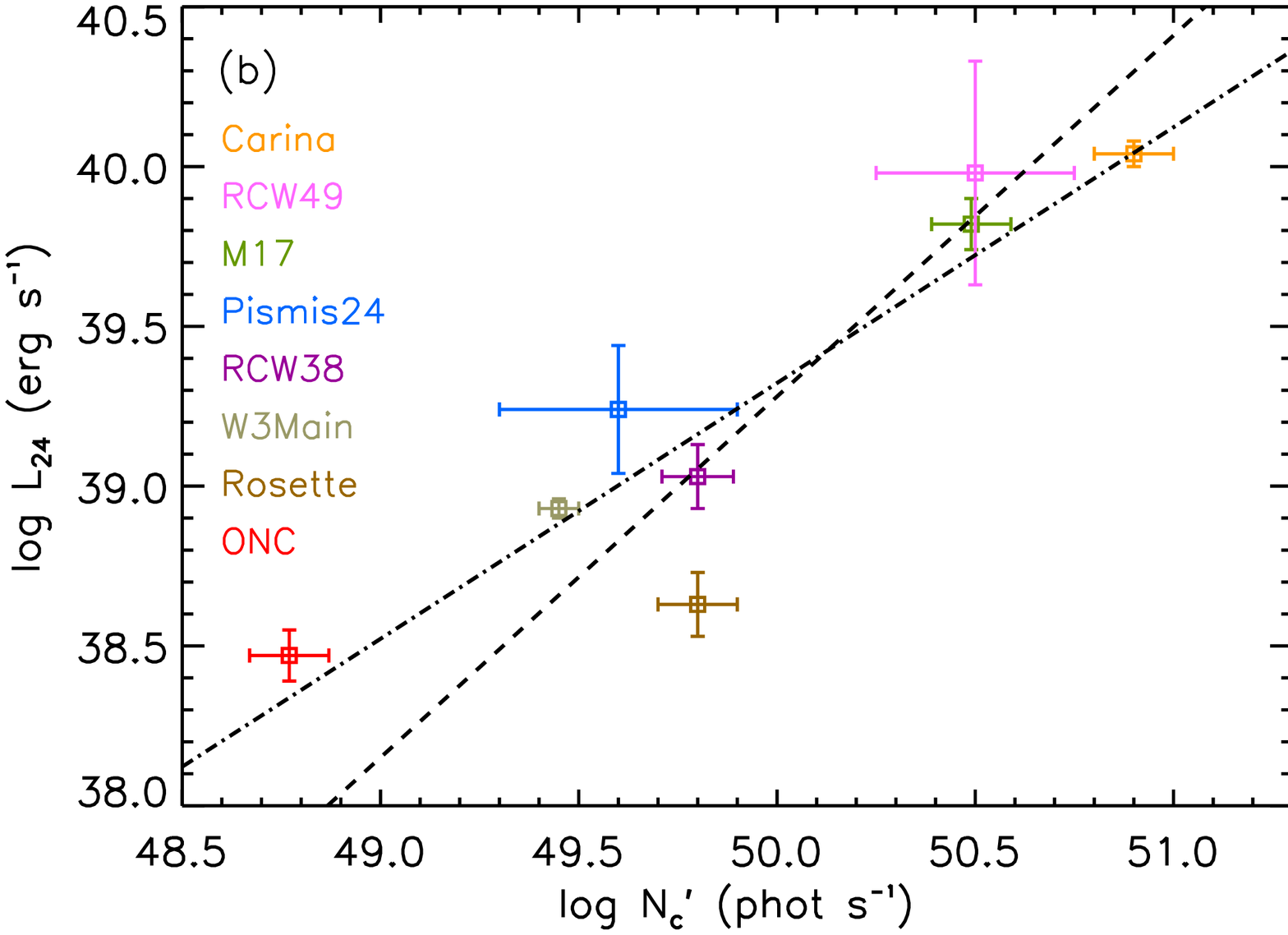}
\plottwo{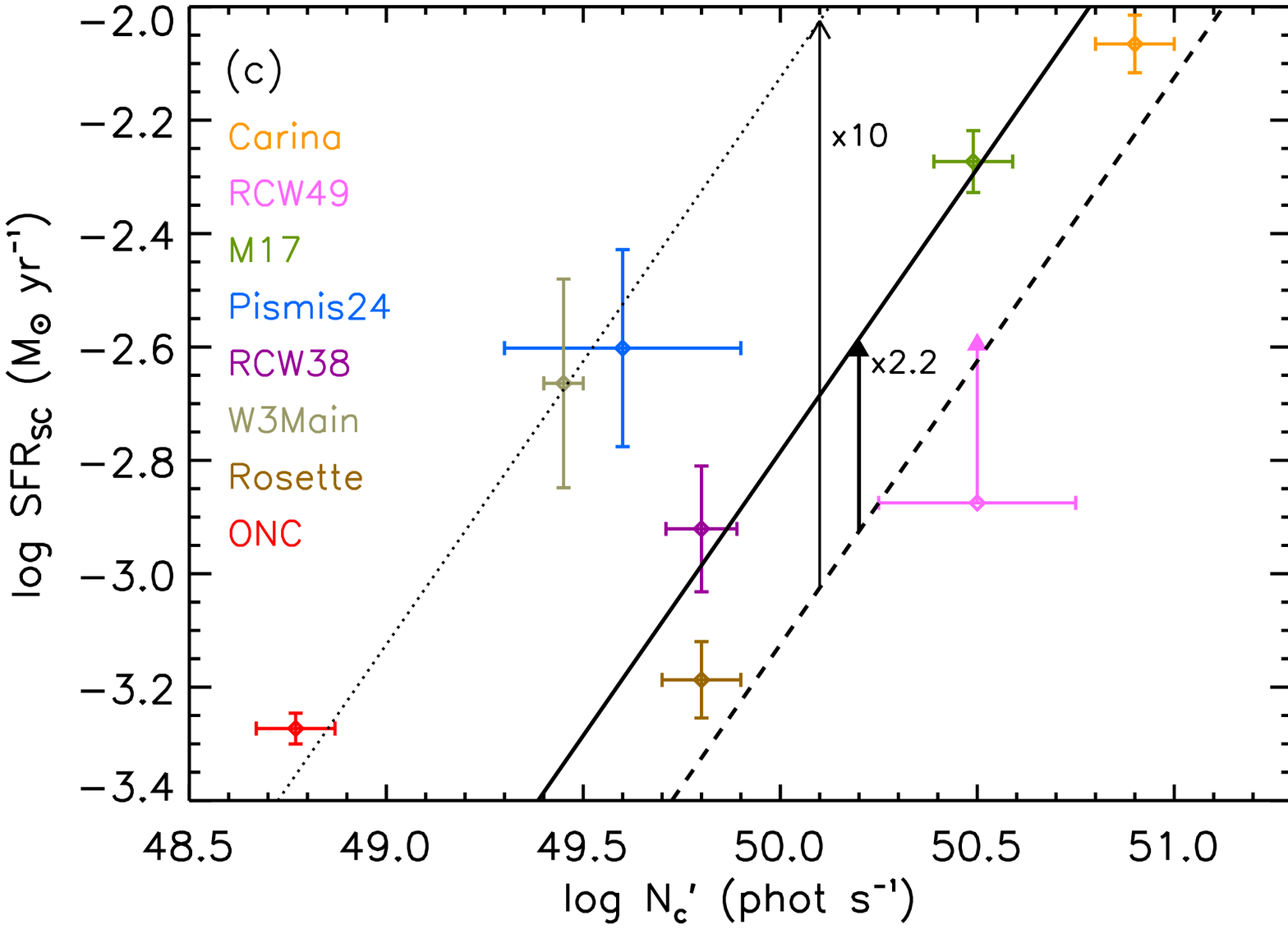}{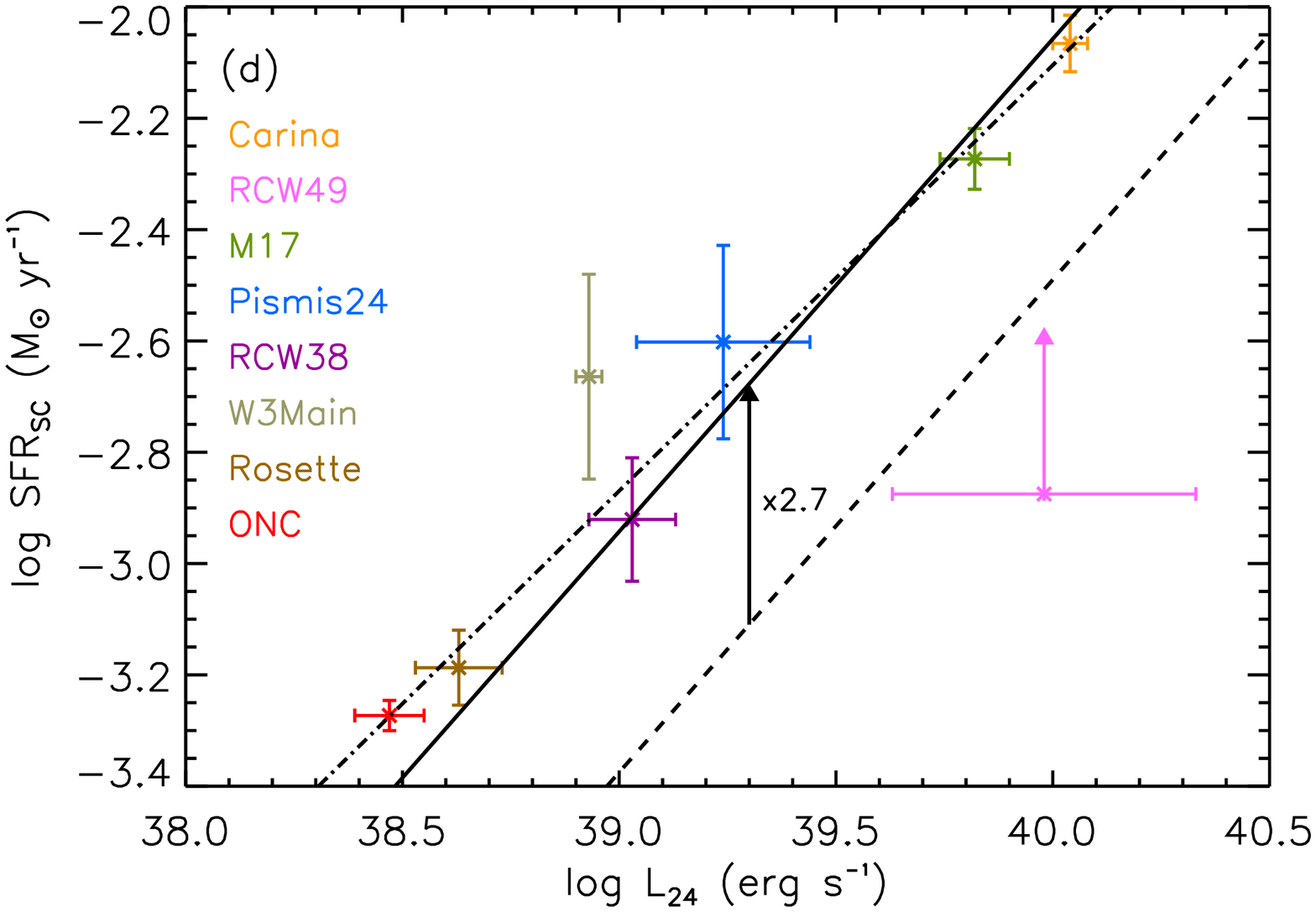}
\caption{({\it a}) Mid-IR SEDs (dots) for 8 Galactic \hii\ regions, from archival {\it IRAS} and {\it MSX} data. Solid curves are spline interpolations to the SEDs. Effective luminosities in the {\it Spitzer}/MIPS 24~\um\ bandpass ($L_{24}$; between the vertical dash-dotted lines) are plotted as boxes. ({\it b}) Plot of $\log{L_{24}}$ versus $\log{N_c'}$, the uncorrected ionizing photon rate. The best fit to the data is plotted as a dash-dotted line with slope $0.80\pm 0.06$. C+07 derived a super-linear relation (dashed line; slope 1.13+/-0.03) for their extragalactic data. ({\it c}) SFRs derived from star counts, $\log{{\rm SFR}_{\rm SC}}$, plotted against $\log{N_c'}$. The relation given in Equation 4, a standard extragalactic calibration, is marked as a dashed line. Versions of Equation 4 scaled up by factors of 2.2 (fitting the weighted mean of the data, excluding RCW 49 and the ONC) and 10 are plotted as solid and dotted lines, respectively. ({\it d}) Plot of $\log{{\rm SFR}_{\rm SC}}$ versus $L_{24}$. The correlation between $\log{{\rm SFR}_{\rm SC}}$ and $\log{L_{24}}$ is consistent with the power-law relation derived by C+07 (dashed line) scaled up by a factor of 2.7 (solid line). The best-fit relation to our data is marked as a dash-dotted line and has a slope $0.76 \pm 0.05$. \label{SFRcal} }
\end{figure*}

{\it Mid-IR luminosity.} Following \citet{Povich_etal07}, we constructed IR SEDs for each \hii~region (Figure~\ref{SFRcal}{\it a}), using aperture photometry to extract background-subtracted flux densities from archival {\it IRAS} \citep{IRAS} and {\it Midcourse Space Experiment} \citep[{\it MSX};][]{Price_etal01} survey images. We then performed a spline interpolation of the SED points, measured the equivalent 24~\um\ flux density $F_\nu(24)$ as the mean value of the interpolation within the MIPS 24~\um\ bandpass, and calculated $L_{24}$ using Equation~\ref{L24}. The dominant source of uncertainty in $L_{24}$ is the uncertain distance to each region (Table~\ref{G8}).

{\it Ionizing photon rates from radio continuum emission.}
We calculated $N_c'$ from the 5 GHz thermal radio continuum flux of each \hii~region using the same conversion as \citet{Smith_etal78}. Because these regions are large on the sky, most of the integrated flux measurements come from single-dish observations that are decades old. \citet{Paladini_etal03} compiled data from  several large surveys of Galactic radio \hii~regions, and we use these cataloged flux densities for all regions except the Carina Nebula and the Rosette, for which we adopt the values of \citet{Smith_Brooks07}, and \citet{Celnik85}, respectively.
The available radio observations form a highly inhomogeneous dataset, often suffering from background contamination and potentially unreliable calibration. We make no attempt to correct for internal absorption of ionizing photons and assume, as we found for the case of M17, that $N_c\approx N_c'$.

In Figure~\ref{SFRcal}{\it b} we plot $\log{L_{24}}$ against $\log{N_c'}$, the relation that provides the foundation for the C+07 extragalactic mid-IR calibration, for which $N_c$ is a direct proxy for SFR. We find a fairly tight correlation (rms scatter of 0.24 dex) with a marginally sub-linear slope ($\log{L_{24}} \propto [0.80\pm0.06]\log{N_c'}$; dash-dotted line). The most significant outlier is the Rosette, which is the only ``extended low-surface brightness'' radio \hii\ region in the sample \citep{Celnik85}, 
hence it is possible that its measured radio continuum flux density includes a significantly larger contribution from faint, extended emission compared to the other regions.
C+07 derived a super-linear relation ($\log{L_{24}} \propto [1.13\pm 0.03]\log{N_c'}$; dashed line) for their extragalactic relation, and suggested that a linear relation could apply to low-luminosity \hii\ regions. Our data would be consistent with a linear relation if the ONC were excluded from the sample.

In Figure~\ref{SFRcal}{\it c} we compare SFR$_{\rm SC}$ with $N_c'$.
While \hii\ regions with higher SFR$_{\rm SC}$ also trend toward higher $N_c'$, all points (except for RCW 49 for which SFR$_{\rm SC}$ is a lower limit only) lie above the linear correlation expected from the extragalactic calibration (Equation \ref{eq:nlyc}; plotted as a dashed line), some by as much as a factor of ${\sim}10$ (dotted line). The weighted mean offset of the data (excluding RCW 49 and the ONC) from the predicted correlation is a factor of 2.2 (solid line), with very large scatter (0.38 dex, rms). Note that the error bars on SFR$_{\rm SC}$ reflect only the uncertainty in $M_{\rm pop}$, which is {\it not} independent of the uncertainty in $N_c'$ (or $L_{24}$ in Panel {\it d}) because both incorporate the distance uncertainty. 
The systematic offset from Equation \ref{eq:nlyc} might be explained in part by correcting $N_c'$ for absorption by dust and escape of ionizing photons, but this correction is unlikely to shift any points by more than a factor of $\sim$2, and as we showed for M17, some points would not shift at all. 

In Figure~\ref{SFRcal}{\it d}, ${\rm SFR}_{\rm SC}$ is plotted against $L_{24}$; this correlation is significantly tighter than the above comparisons of ${\rm SFR}_{\rm SC}$ and $N_c'$, with an rms scatter of 0.11 dex. The best-fit (dash-dotted; excludes RCW 49) line has slope $0.76\pm 0.05$, shallower than the 0.8850 slope of the analogous C+07 extragalactic relation (dashed line). The C+07 calibration, scaled up by a factor of $2.7$ (solid line), best matches the weighted mean correlation between SFR$_{\rm SC}$ and $L_{24}$.
Because our estimates of SFR$_{\rm SC}$ (Table~\ref{G8}) are lower limits, the factor of 2.7 represents a {\it minimum} systematic discrepancy between the mid-IR SFR calibrations and SFR measured from the resolved stellar populations.

There are several possible explanations for the decreased scatter in the ${\rm SFR}_{\rm SC}$--L$_{24}$ correlation, as compared with the ${\rm SFR}_{\rm SC}-N_c'$ correlation, including region-to-region variations in the correction factor for $N_c'$ to $N_c$, inhomogeneity of radio observations, and/or stochastic sampling of the high-mass tail of the IMF.
The last possibility arises because ionizing photon rate produced by individual stars is a steeper function of mass than the bolometric luminosity, hence in extreme cases only the single most massive star in a given region dominates $N_c$, while a wider range of stellar masses can heat dust and contribute to the 24 $\mu$m emission. If the mass of the most massive star is vulnerable to large, stochastic fluctuations from cluster to cluster, we might expect $L_{24}$ to trace SFR more smoothly and with less scatter than $N_c$.

In summary, we find that SFR$_{\rm SC}$ derived from low- and intermediate-mass star counts to be systematically higher, by factors of at least 2--3, compared to either the calibration of Equation \ref{eq:nlyc} (Figure~\ref{SFRcal}{\it c}) or the C+07 extragalactic SFR--$L_{24}$ calibration  (Figure~\ref{SFRcal}{\it d}).
The discrepancy between SFR$_{\rm SC}$ and SFR$_{24}$ (or equivalently SFR$_{ff}$) we derived for the specific case of M17 (\S\ref{m17ir}) therefore appears to be a general feature of Galactic \hii~regions.
Parallels might be drawn between these results and the findings of \citet{Heiderman_etal10}, who reported SFR surface densities measured from YSO counts in low-mass, Galactic molecular clouds that are higher by factors of 5--10 compared to the standard extragalactic Schmidt--Kennicutt Law \citep{Kennicutt98b}.

\subsection{Systematics Affecting SFR Calibrations Based on Lyman Continuum Photon Rates} \label{sfrdiff}

From the analysis in the preceding subsections, we conclude that  even if Lyman continuum photon rates were known to arbitrarily high precision, an intrinsic, systematic uncertainty in the derived SFRs would remain. Specifically, it appears that the mid-IR SFR calibration of C+07 and, by extension, other calibrations that rely upon Lyman continuum photon rates {\it underestimate} SFRs by at least a factor of 2--3 (Figures~\ref{SFRcal}{\it c,d}). This systematic affects the coefficient in Equation \ref{eq:nlyc}, the transformation between SFR and Lyman continuum photon production rate. The coefficient can be parameterized as $M_{\rm pop}/(\tau N_c)$, where $M_{\rm pop}/N_c$ is the mass of a young stellar population normalized by its production rate of Lyman continuum photons, and the timescale $\tau$ is either the duration of star formation traced by the observations of $N_c$ or the lifetime of the ionizing stars, whichever is shorter. $M_{\rm pop}/N_c$ is itself a function of age in population synthesis models (e.g., the Starburst99 models; \citealp{Vazquez_Leitherer05}), but its functional form cannot be realistically modeled for cases where the SFR varies on timescales shorter than O-star lifetimes. As we cautioned in \S2, Equation 4 strictly describes the SFR required to maintain a steady-state population of ionizing stars, implicitly assuming $\tau \approx 5$ Myr.
Several effects likely to be important to $M_{\rm pop}/N_c$ and one important consequence of the assumed $\tau$ are discussed below.

\subsubsection{Uncertain Lyman Continuum Photon Production Rates for O Stars}
Fundamentally, $M_{\rm pop}/N_c = (1/\eta)\sum_{i}(m/Q_0)_{i}$, where $m/Q_0$ is the ratio of stellar mass to Lyman continuum photon production for individual O stars and $\eta$ is the (small) fraction of $M_{\rm pop}$ contained in O stars according to the assumed IMF. Values for $Q_{0}$ must be obtained from models of O stars as functions of spectral type and luminosity class. The Starburst99 grid of population synthesis models \citep{Vazquez_Leitherer05} incorporates the modern O star calibrations and synthetic atmospheres of \citet{Smith_etal02} and \citet{Martins_etal05}.  These state-of-the-art calibrations employ non-LTE stellar dynamic atmosphere models that  include the effects of line-blanketed winds and predict cooler effective temperatures by as much as ${\sim}5000$~K for the earliest spectral types. This results in ${\sim}40\%$ reductions of $Q_0$ for a given spectral type compared to earlier models \citep{Panagia73, Vacca_etal96}. Despite the obvious improvements in models of massive stars, we reiterate the caveat of \citet{Martins_etal05}: the uncertainty on $Q_0$ remains a factor of ${\sim}2$ in any model, hence its impact cannot be neglected, although it is unclear in which direction this uncertainty might bias SFR determinations.

\subsubsection{The O star mass discrepancy }
The uncertainty in the numerator of $m/Q_0$ is also important. Model-based calibrations of O star parameters predict the stellar mass as a function of spectral type in two different ways. The emergent spectrum gives values for stellar luminosity $L$, effective temperature $T_{\rm eff}$, and effective surface gravity $g_{\rm eff}$ measured from spectral line broadening, giving
\begin{equation}
  M_{\rm spec} = \frac{g_{\rm eff}L}{4\pi G\sigma T^4_{\rm eff}},
\end{equation}
\citep{Vacca_etal96, Martins_etal05} where $\sigma$ is the Stefan-Boltzmann constant and $G$ is the gravitational constant. Alternatively, $L$ and $T_{\rm eff}$ from stellar evolutionary tracks can be used to determine an evolutionary mass $M_{\rm evol}$, which is associated with a stellar gravity $g_{\rm evol}$ \citep{Vacca_etal96}. In general, $g_{\rm eff}<g_{\rm evol}$, because a strong stellar wind lifts the surface layers of a star, reducing the gravity probed by spectral lines formed in the photosphere. Consequently, there is a discrepancy in the derived masses of up to a factor of ${\sim}2$, a systematic effect with $M_{\rm spec}<M_{\rm evol}$. O star masses based on $M_{\rm spec}$ in stellar models could therefore be underestimates, particularly for the earliest spectral types \citep{Martins_etal05}.
This would tend to bias $m/Q_0$ downward in population synthesis models, leading to underestimates of $M_{\rm pop}/N_c$ and hence SFRs. \citet{Weidner_Vink10} recently found good agreement between $M_{\rm  spec}$ and dynamical masses calibrated to massive eclipsing binary systems, so it is possible that the O star mass discrepancy has been resolved.

\subsubsection{The Upper Mass Limit of the IMF}
The most massive stars ($m\ge 40$~\Msun) dominate $N_{c}$ but represent only $\eta=0.07$ of $M_{\rm pop}$ (assuming the Kroupa IMF) and follow, very approximately, the relation $Q_0(m)\propto m^{3.5}$. The ratio $M_{\rm pop}/N_{c}$ is therefore sensitive to the assumed upper mass limit of the IMF. This cutoff is poorly known; for example, \cite{Zinnecker_Yorke07} conclude that an upper limit of 150 \Msun\ is consistent with data, while \cite{Crowther_etal10} claim a maximum stellar mass of 300 \Msun. This confusion may be compounded by mounting evidence that many O stars form in binary systems with nearly equal-mass components \citep{Sana_etal08}. \citet{MA08} performed numerical experiments testing the effect on observational IMF determinations of mistaking unresolved massive binaries for single stars and concluded that while neglecting stellar multiplicity has a negligible effect on the high-mass IMF slope, the high-mass cutoff of the IMF will be systematically overestimated. In this work, we have conservatively assumed an upper mass limit of 100 \Msun\ (Equation~\ref{eq:hi-mIMF}). Increasing this limit would decrease the SFR estimated from a given rate of ionizing photons, only worsening the discrepancy between SFR$_{24}$ and SFR$_{\rm SC}$.

\subsubsection{Runaway Massive Stars} 
Studies of massive stars in the field (remote from any significant episode of star formation) show that $\sim$10--30\% of O stars escape their natal star-forming regions \citep{Gies87, Moffat_etal98, Oey_Lamb11}. For young \hii\ regions, O stars are likely ejected via dynamical encounters, which are enhanced by their high binary fraction \citep[e.g.,][]{Leonard_Duncan90, Kroupa04}; direct signatures of O-star escape can be seen in imaging of bow shocks \citep{Gvaramadze_etal11}. Although the loss of ionizing stars from \hii\ regions could contribute to the discrepancy between SFR$_{24}$ and SFR$_{\rm SC}$, it can not account for the full magnitude of the discrepancy.

\subsubsection{The Assumed Form of the IMF}

Fundamentally, the discrepancy between SFR$_{\rm SC}$ and SFR$_{24}$ may indicate that $M_{\rm pop}/N_{c}$ predicted by stellar population synthesis models is systematically too low or, equivalently, the fraction of mass $\eta$ contained in massive stars is too high. A straightforward way to increase $M_{\rm pop}/N_c$ would be to increase the numbers of low-mass stars (which contain the bulk of the stellar mass) relative to massive stars (which dominate the luminosity in young populations). This is equivalent to steepening the intermediate-mass (or super-solar) IMF slope. In their recent review, \citet{Bastian_etal10} argue that there is no convincing evidence for strong IMF variations, but this conclusion does not amount to a claim that the form of the ``universal'' IMF is known to high precision.
As is the case with SFRs, IMF {\it comparisons} across different environments are more reliable than {\it absolute} measurements. Indeed, Figure~2 of \citet{Bastian_etal10} reveals that while derivations of $\Gamma$ in the super-solar mass regime are broadly consistent with the Salpeter--Kroupa slope ($\Gamma=1.3$), this actually represents only a lower bound on $\Gamma$ reported for 1.5~\Msun$<m<5$~\Msun. At intermediate masses, the IMF slopes cluster around $\Gamma'=1.7$ \citep{Scalo98}. 

If we alter our standard Kroupa IMF (Equations 1--3) by increasing $\Gamma = 1.3$ to $\Gamma' = 1.7$ for 1.5~\Msun$<m<5$~\Msun, we derive SFR$'_{24}=\eta/\eta'\times {\rm SFR}_{24}=1.4\times {\rm SFR}_{24}$. We also find SFR$'_{\rm SC}=0.9\times {\rm SFR}_{\rm SC}$, by normalizing both IMFs to the same $N_{\rm pop}$. A steeper intermediate-mass IMF slope marginally reduces SFRs based on low/intermediate-mass star counts while significantly increasing SFRs derived from massive star diagnostics, driving the independent measurements toward convergence. The combined effect produces a factor of ${\sim}1.6~(=1.4/0.9$) increase in ${\rm SFR}_{24}/{\rm SFR}_{\rm SC}$, enough to explain much of the systematic discrepancy.

It would be reasonable to question the appropriateness of altering the IMF used in the XLF scaling method, given that a fundamental assumption of this method is similarity with the ONC IMF (\S\ref{m17xlf}). Hence we note that the benchmark ONC IMF of \citet{Muench_etal02} was measured using low-mass stars only ($m<2$~\Msun; see Figure~\ref{IMF_cartoon}), and $\Gamma=1.21$ was simply extrapolated to the high-mass tail. In spite of its status as the nearest ``massive'' young cluster, the ONC contains too few intermediate-mass stars to support a statistical measurement of the super-solar IMF slope. We conclude that a super-solar IMF slope steeper than Salpeter is not excluded by existing IMF determinations, and our method for comparing SFRs derived from XLF scaling and massive star tracers can indirectly constrain its value.

\subsubsection{Stochastic Sampling of the IMF}

In the limit of very low SFR, the linear calibration of SFR against ionizing photon rate (Equation~\ref{eq:nlyc}) must break down, as star formation can occur without producing any massive stars in a given region. In the case of most Galactic \hii\ regions, we must consider the intermediate possibility that the high-mass tail of the IMF will not be fully sampled. To representatively sample the high-mass tail of the IMF, a stellar population should contain $\gtrsim$ 10 O stars \citep{Cervino_Valls-Gabaud03}, which corresponds to $M_{\rm pop}=2,900$ M$_{\odot}$ (assuming a Kroupa or Chabrier IMF; \citealt{Lee_etal09}). \citet{Cervino_etal03} also find that a total stellar mass of ${\sim}3,000$ M$_{\odot}$ is the lowest mass for which sampling effects can be ignored (for a young stellar population and solar metallicity). Similarly, \citet{DaSilva_etal11} find that, after accounting for stochastic effects, Lyman continuum fluxes are consistent on average with those expected for a fully sampled IMF for SFRs $\gtrsim 10^{-3}$ M$_{\odot}$ yr$^{-1}$ (although the scatter in $N_c$/SFR is significant for SFR $= 10^{-3}$ M$_{\odot}$ yr$^{-1}$, approximately a factor of two). For SFR $= 10^{-4}$ M$_{\odot}$ yr$^{-1}$, the average ionizing luminosity is suppressed by a factor of $\sim$2.

The eight Galactic \hii\ regions described in Table \ref{G8} span a wide range of $M_{\rm pop}$ and SFRs, with some significantly above these limits (Carina, M17) and others well below them (ONC, Rosette). In the lowest-mass regions, we might therefore expect calibrations based on population synthesis models with fully-sampled IMFs to underestimate SFR. However, stochastic sampling cannot explain the systematic offset of {\it all} \hii\ regions in our sample, including the high-mass regions Carina and M17 that fully sample the IMF, from the C+07 relation. We note that the best-fit line to the SFR$_{\rm SC}$--$L_{24}$ relation (Figure~\ref{SFRcal}{\it d}) has a shallower slope than the C+07 correlation, as would be expected if $L_{24}$ preferentially underestimates SFRs in the lower-mass regions, such as the ONC.


\subsubsection{Star Formation Timescales} \label{ages}

As shown in Table~\ref{G8}, the (upper limit) ages for our Galactic \hii~region sample range from 1 to 5~Myr, with a median age of 2~Myr. These young ages are a consequence of our selection criterion (II), that most of the bolometric luminosity be reprocessed by warm dust and reradiated in the mid-IR. Older Galactic \hii~regions have been excluded from our analysis because they do {\it not} exhibit bright mid-IR nebulosity and would not be identifiable as 
24~\um\ point sources in a SINGS image of an external galaxy, hence they would not have been included in the C+07 calibration sample. Prominent excluded clusters include Westerlund 1, one of the most massive young clusters in the Milky Way \citep{Clark_etal05}, and several Red Supergiant Clusters \citep{Figer_etal06,Davies_etal07,Alexander_etal09}, all of which contain significant populations of Wolf-Rayet stars, yellow hypergiants, red supergiants, and other evolved massive stars. The ages of these clusters fall in the 5--20~Myr range, hence they represent the future evolution of the giant \hii\ regions in our sample. The dust in such evolved \hii\ regions has either been destroyed by the hard radiation field or blown away by stellar winds and possibly supernovae, removing the source of the bright mid-IR emission. Such regions certainly contribute to the extended IR emission of the Galaxy, since very little of their bolometric luminosity is reprocessed locally by dust. The integrated Galactic IR emission is dominated by cool dust with a significant component coming from older stars \citep[e.g.][]{Misiriotis_etal06}, whereas the SEDs of young Galactic \hii~regions are dominated by warm dust heated by O stars (Figure~\ref{SFRcal}{\it a}). 

The median value for the timescale $\tau_{\rm SF}$ used to calculate SFR$_{\rm SC}$ is a factor of $\sim$2.5 times {\it lower} than O-star lifetimes, the timescale assumed implicitly by the C+07 calibration and other calibrations based on population synthesis models, like Equation~\ref{eq:nlyc} (see caveat in \S\ref{inputs}). Hence the discrepancy between SFR$_{\rm SC}$ and $L_{24}$ (or $N_c'$) in very young regions could be resolved by applying this correction factor
to the calibrations of SFR versus diffuse emission tracers.

The C+07 \hii~region sample was selected on the basis of compact, bright ``knots'' of mid-IR emission. Based on the discussion above, this selection criterion likely imposed an age constraint of $\tau_{\rm SF}\la 5$~Myr, which means that {\it absolute} SFRs derived by C+07 and similar studies of individual extragalactic star-forming regions could be systematically underestimated by factors of 
${\ga}8~{\rm Myr}/5~{\rm Myr}=1.6$, where 8 Myr is the steady-state timescale assumed implicitly by population synthesis (Section 2) and 5 Myr is the {\it maximum} age of a mid-IR bright \hii~region. We note that the \hii\ region models employed by C+07 neglected \hii\ region expansion, stellar winds, and supernovae, without which \hii\ regions could remain mid-IR bright throughout the lifetime of the ionizing stars. This discrepancy due to assumed timescales should not affect studies like K+09, where SFRs are measured as averages over entire galaxies (although care is needed in the case of dwarf galaxies and other systems subject to short bursts of star formation). In the case of normal spiral galaxies like the Milky Way, it is usually a good approximation that the galaxy-wide SFR has been constant over the last ${\sim}$100 Myr, and the calibration of Equation~\ref{eq:nlyc} should hold. Indeed, K+09 find that IR traces SFR more weakly in their data than in the C+07 sample, consistent with the longer timescales traced by galaxy-wide H$\alpha$ and IR emission.


\section{Are Global Galactic SFR estimates consistent with extragalactic diagnostics?} \label{extest}

As we have seen, SFR estimates depend on a wide range of assumptions; incorrect input assumptions can lead to potentially large systematic errors in absolute SFR determinations. In addition, we must keep in mind that different assumptions may be appropriate in different observational regimes, complicating attempts to compare SFRs. For example, if one set of assumptions were 
valid in external galaxies (which are often measured as wholes) while a different set of assumptions held in the Milky Way (where star formation is often studied piecemeal, and always in projection), these inputs could lead to a discrepancy between SFRs in the Milky Way and other galaxies.

As reviewed in \S\ref{mwdiag}, significant effort has been invested in measuring the Galactic SFR, but if we were external observers located Mpc from the Galaxy, would we still measure a Milky Way SFR of $\sim$2~M$_{\odot}$~yr$^{-1}$? The answer to this question is critical for comparing the Milky Way with other galaxies and extrapolating the detailed knowledge we gain in our Galaxy to more distant systems. 
Here we present two tests that can assess, directly and with a minimum of assumptions, whether the Galactic SFR is consistent with extragalactic SFR calibrations. 
Although preliminary, 
these tests can serve as valuable sanity checks for our comparison of the Galactic SFR with calibrations used in other galaxies.

\subsection{The Brightest Radio Supernova Remnant (Cassiopeia A)} \label{casa}

The luminosity functions (LFs) of radio SNRs have been observed in $\sim$20 galaxies and can be modeled as power laws with constant index and scaling that depends roughly linearly on SFR (\citealt{Chomiuk_Wilcots09b}; hereafter CW09). Unfortunately, it is very difficult to establish a complete sample of SNRs in the Milky Way \citep[e.g.,][]{Green05}, so a comparison of the full LF is not a plausible technique for estimating the Galactic SFR. However, CW09 showed that there is a good correlation between a galaxy's SFR and the radio spectral luminosity of its most radio-bright SNR ($L_{\rm max}$). We can use this $L_{\rm max}$--SFR correlation to make an independent estimate of the Galactic SFR, assuming that Cas A is the most luminous SNR in the Milky Way. This is a valid assumption because any SNR with a similar luminosity to Cas A would be bright enough for inclusion in any modern catalog, even if it were located on the far side of the Galaxy \citep{Green04}. The details of our extragalactic sample are described in the Appendix; SFRs are measured using the K+09 calibration of H$\alpha$ and total IR emission. Here we measure SNR luminosities at 4.85 GHz because this frequency is less susceptible to the free-free absorption that can dampen lower-frequency radio continuum emission in starburst galaxies.


\begin{figure}[t]
\epsscale{1.0}
\plotone{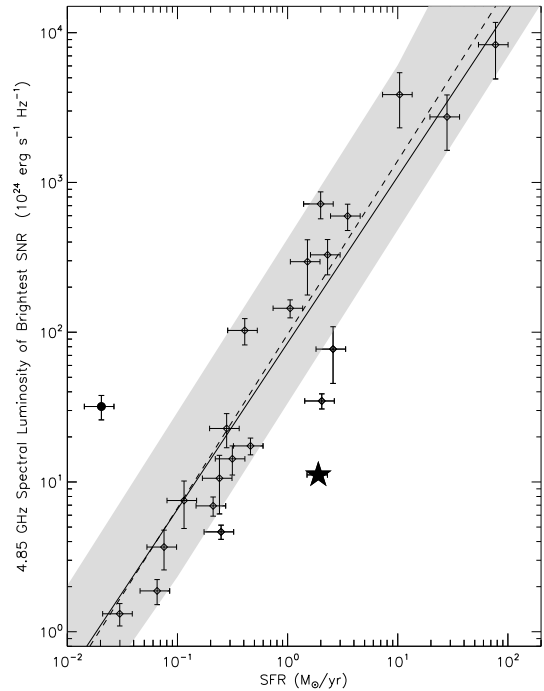}
\caption{For each galaxy in our sample, we plot the 4.85 GHz radio luminosity of its brightest SNR as a function of the galaxy's SFR. The solid line marks the best fit to the data. The dashed line represents the expected correlation due to random statistical sampling of a power-law LF, and the grey shaded region shows the expected 1$\sigma$ variation in the correlation due to stochasticity. The point marked by a filled circle is the outlier IC 10. The Milky Way is represented by a black star, assuming the Milky Way's most luminous SNR is Cas A and its SFR is $1.9 \pm 0.4$ M$_{\odot}$ yr$^{-1}$.}
\label{sfr_maxlum5}
\end{figure}

There is a clear correlation between $L_{\rm max}$ and SFR (Figure \ref{sfr_maxlum5}), best fit at 4.85 GHz as: 
\begin{equation}
L_{\rm max}^{4.85} = \left(85^{+22}_{-17}\right) \textrm{SFR}^{1.11\pm0.15}
\end{equation}
This fit is shown as a solid line in Figure \ref{sfr_maxlum5}. This correlation is consistent with stochastic sampling of the power-law SNR LF (measured as described in the Appendix and CW09), marked as a dashed line in Figure \ref{sfr_maxlum5}. The scatter in the $L_{\rm max}$--SFR correlation is $\pm$0.51 dex, consistent with the scatter expected from stochastic sampling of a power law ($\pm$0.55 dex).  

 We note that IC 10 is a clear outlier (marked as a filled circle in Figure \ref{sfr_maxlum5}); its large ``superbubble" \citep[$\gtrsim$ 130 pc][]{Yang_Skillman93} is much more luminous than expected for a galaxy with IC 10's SFR. \cite{Yang_Skillman93} claim that the most likely explanation for this source is multiple SNRs overlapping in the vicinity of a giant \ion{H}{2} region. In fact, many of the brightest SNRs in galaxies may not be typical of the SNR population as a whole, and they are likely to be a rather heterogeneous class of objects \citep{Chu_etal99}. Some may be the combined emission from several overlapping SNRs \citep{Yang_Skillman93} or the remnants of hypernovae/gamma-ray bursts \citep{Lozinskaya_Moiseev07}. Others may be interacting with unusually dense circumstellar material blown off by the progenitor star \citep[e.g., the luminous SNR in NGC 4449;][]{Milisavljevic_Fesen08}. Others may not be SNRs at all---\cite{Pakull_Grise08} point out that many large-diameter (200--500 pc) bright radio bubbles are likely blown by the jets of ultraluminous X-ray sources (ULXs) and not by SNe. For example, in the most luminous radio ``SNR" in NGC 7793 \cite[S26;][]{Pannuti_etal02}, \cite{Pakull_Grise08} uncover a jet-like structure via high-resolution X-ray imaging which spans the $\sim$450 pc radio bubble and implies that that this bubble is blown by a microquasar and not by SN activity. Regardless of the physical mechanisms driving the $L_{\rm max}$--SFR relation, 
it appears to hold empirically across more than three orders of magnitude, and we can use it as a test of current estimates of the Galactic SFR.

In Figure \ref{sfr_maxlum5}, we plot the luminosity of Cas A and a Galactic SFR of $1.9\pm0.4$ M$_{\odot}$ yr$^{-1}$ (as found in \S3), marked with a solid star. The Milky Way appears to deviate from the $L_{\rm max}$--SFR relation as the most outlying point besides IC 10. Using the technique described in the Appendix, we calculate a 0.2\% probability that, given the luminosity of Cas A, the SFR of the Milky Way is $1.9\pm0.4$ M$_{\odot}$ yr$^{-1}$ (or, a 3\% probability that it it $>$ 1 M$_{\odot}$ yr$^{-1}$). Given the inherent uncertainty in the $L_{\rm max}$--SFR relation and the fact that outliers like IC 10 do exist, we can not conclusively eliminate the possibility that Galactic SFR estimates and extragalactic SFR diagnostics are consistent with one another. However, it is worth noting that the observed offset of the Milky Way from the $L_{\rm max}$--SFR relation implies that either the Galactic SFR has been overestimated or that extragalactic SFRs have been systematically underestimated. 

\subsection{Luminous Young X-ray Point Sources} 

We can further test if the measured Galactic SFR is consistent with SFR determinations in other galaxies using young X-ray point sources (YXPs). The X-ray emission from galaxies has often been claimed to trace SFR \citep[e.g.,][]{Grimm_etal03, Ranalli_etal03, Mineo_etal11}, because of the high luminosities from high-mass X-ray binaries (HMXBs), in addition to SNRs and ULXs.


\begin{figure}[htp]
\centering
\includegraphics[width=11cm, angle=90]{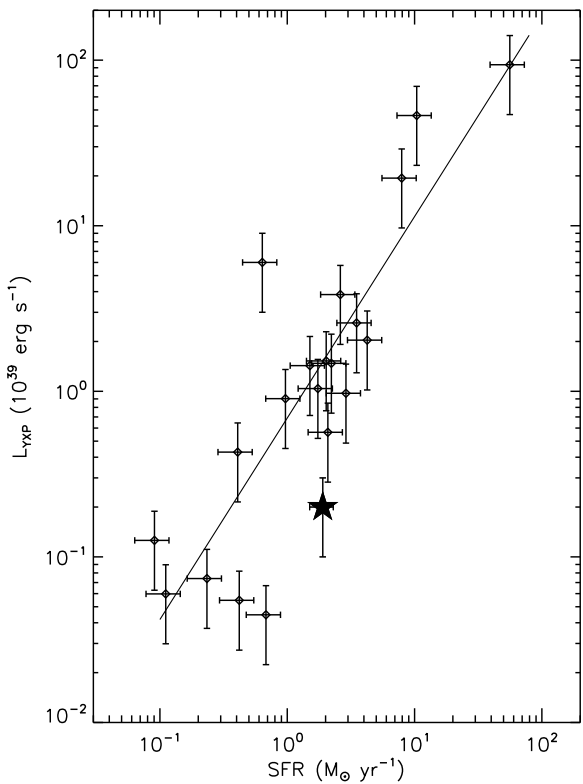}
\caption{The integrated 2--10 keV luminosity of young X-ray point sources in each of 20 nearby galaxies \citep{Persic_Rephaeli07} plotted against SFR (assuming 50\% uncertainty on $L_{\rm YXP}$). The best fit correlation is marked as a solid black line. The Milky Way is plotted as a black star, assuming the SFR from \S3 of $1.9\pm0.4$ M$_{\odot}$ yr$^{-1}$ and HMXB luminosity from \cite{Grimm_etal02}.}
\label{sfr_xlum}
\end{figure}

\cite{Grimm_etal02} studied what the Milky Way X-ray source population would look like to an outside observer and estimated a total luminosity of $L_{\rm HMXB} = (2\pm1)\times10^{38}$~erg~s$^{-1}$ over 2--10~keV for HMXBs. We also add the luminosity of the Crab Nebula ($L_{\rm X} = 1 \times 10^{38}$ erg s$^{-1}$ assuming a distance of 2 kpc; \citealt{Grimm_etal02, Davidson_Fesen85}), which dominates the Milky Way's SNR luminosity in the X-ray band. The Milky Way does not host any ULXs, so its $L_{\rm YXP} = (3\pm1)\times10^{38}$~erg~s$^{-1}$.

We can compare the Milky Way with the sample of 20 ``local normal and starburst'' galaxies in \cite{Persic_Rephaeli07}, for which there are X-ray point-source LFs in the literature. Persic \& Rephaeli acknowledge that the major uncertainty in using the X-ray point-source luminosity as a SFR indicator is contamination by low-mass X-ray binaries (LMXBs), which are associated with older stellar populations and trace stellar mass rather than SFR. In the case of the Milky Way, the integrated luminosity of LMXBs is ten times greater than that of HMXBs; \citealt{Grimm_etal02}).  Fortunately, the relative number of HMXBs as compared with LMXBs should be a function of SFR divided by a galaxy's stellar mass \citep{Gilfanov_etal04}. \cite{Persic_Rephaeli07} deal with LMXB contamination by assuming that X-ray point-source LFs in systems with high SFR are representative of the young component and universally applicable; similarly, they assume that LFs measured in E/S0 galaxies represent the old population of X-ray point sources. They fit each galaxy's LF with a combination of old and young components. In Fig. \ref{sfr_xlum}, we plot the product of their measured point-source luminosities and YXP fractions, rescaled to the distances listed in \cite{Kennicutt_etal08} and \cite{Moustakas_Kennicutt06}. We calculate SFRs as described in the Appendix, using the K+09 calibration.

In Fig. \ref{sfr_xlum}, we see that there is indeed a correlation between $L_{\rm YXP}$ and SFR. The Milky Way, assuming a SFR of $1.9\pm0.4$ M$_{\odot}$ yr$^{-1}$, is marked as a star. The standard deviation around the best fit line is 0.47 dex, and the Milky Way is located $\sim$1.5$\sigma$ from the correlation. The Milky Way's deviation corresponds to an unusually high SFR for its measured $L_{\rm YXP}$, an offset in the same direction as observed for the $L_{\rm max}$--SFR relation. Recently, \citet{Mineo_etal11} pointed out that the Persic \& Rephaeli correlation is offset downwards from other X-ray binary--SFR relations in the literature by a factor of $\sim2-2.5$, fitting unusually low $L_{\rm YXP}$ values as a function of SFR. They hypothesize that \citet{Persic_Rephaeli07} have overestimated the contribution from the old population of X-ray sources; if this is true, it would imply that the discrepancy of the Milky Way from the $L_{\rm YXP}$--SFR relation is larger than that shown in Fig \ref{sfr_xlum}.
In addition, \citet{Lehmer_etal10} show that the Milky Way is a slight ($\sim 1 \sigma$) outlier from their X-ray calibration to external galaxies. They find that given the Milky Way's mass ($5 \times 10^{10}\ \textrm{M}_{\odot}$; \citealt{Hammer_etal07}) and SFR (assumed to be 2.0 M$_{\odot}$ yr$^{-1}$, in good agreement with that used here), the ratio of X-ray luminosity in LMXBs as compared with HMXBs should be $\sim$1.4, almost an order of magnitude lower than the $L_{\rm LMXB} /  L_{\rm HMXB} \approx 10$ measured for the Milky Way by \citet{Grimm_etal02}. This is consistent with a SFR of 2.0 M$_{\odot}$ yr$^{-1}$ for the Milky Way being higher than expected from extragalactic calibrations. 

\subsection{Is the Milky Way an Outlier?}

The Milky Way appears to deviate from both the $L_{\rm YXP}$--SFR and $L_{\rm max}$--SFR relations at the $1-3 \sigma$ level, consistently showing a higher SFR than expected from the extragalactic relations. This deviation is marginal, and the Milky Way's similar offset from both correlations might simply be a coincidence, so we can not eliminate the possibility the the Milky Way's measured SFR is consistent with extragalactic determinations. However, both the SNR and YXP tests hint that a Galactic SFR of 2 M$_{\odot}$ yr$^{-1}$ may be higher than we would measure for the Milky Way if we were external observers using the K+09 SFR calibration. If this is true, there are three possible explanations for the difference: (1) estimates of SFR are biased high in the Milky Way (compared with ``absolute'' SFRs) (2) estimates of SFR are biased low in external galaxies; and/or (3) the Milky Way is a fundamentally outlying system (as suggested by \citealt{Hammer_etal07}). 

If this deviation is real, it is unlikely to be related to the differences between Galactic \hii\ regions and extragalactic calibrations described in \S\ref{g-ex}. Assuming that the most luminous SNR and the YXP population roughly trace the counts of massive stars, 
and understanding that the SFRs used in the $L_{\rm YXP}$--SFR and $L_{\rm max}$--SFR relations are predominately measurements of the Lyman continuum rate, the Milky Way's displacement from the $L_{\rm max}$/$L_{\rm YXP}$--SFR relations implies that available measurements of $N_c'$ and $L_{24}$ {\it overestimate} the Galactic SFR, 
whereas we have found that these quantities {\it underestimate} SFRs in individual Galactic \hii\ regions. 
If the offsets described in \S\ref{g-ex} are due to a poorly constrained (but universal) IMF shape, they will affect the Milky Way and external galaxies equally, while differences due to assumptions about star formation timescales should be irrelevant when we average over entire galaxies (including the Milky Way). 

It is therefore difficult to explain the potential offset of the Milky Way from the $L_{\rm max}$/$L_{\rm YXP}$--SFR relations, but we must keep in mind that the Galactic SFR remains quite uncertain. 
For example, to calculate a total ionizing luminosity for the Milky Way from their catalog of WMAP \hii\ regions, \citet{Murray_Rahman10} had to make at least three significant assumptions, each of which increases the Galactic SFR by $\sim$40--100\% with large uncertainty: 1) they multiply their measured $N_c'$ by two to correct for distant \hii\ regions missing from their catalog; 2) they multiply by 1.5 to account for the diffuse component of the WMAP free-free emission, assuming this diffuse component has the same distribution as the detected \hii\ regions; and 3) they multiply by 1.37 to correct for Lyman continuum escape and absorption by dust (detail given in \citealt{McKee_Williams97}; note that the K+09 extragalactic calibration should also account for dust absorption, because it uses a combination of H$\alpha$ and IR emission). In addition, if 1/3 of the free-free emission has a very diffuse, smooth distribution as \citet{Murray_Rahman10} assert, some of this low-surface brightness emission may go undetected in external galaxies, leading to an underestimate of extragalactic SFRs. Additional strategies for comparison between the Milky Way and external galaxies, along with better statistics on the $L_{\rm max}/L_{\rm YXP}$--SFR correlation, are required to test the provocative implication that the Galactic SFR may be overestimated, and/or extragalactic SFRs may be underestimated, by a factor of a few.

\section{Conclusions} \label{concl}

After normalizing published estimates of the Galactic SFR to Starburst99 population synthesis models and a Kroupa IMF, measurements converge to a Galactic SFR of $1.9\pm 0.4$ M$_{\odot}$ yr$^{-1}$. 

We test to see if SFR estimates in the Milky Way are consistent with extragalactic SFR diagnostics, using luminous SNRs and YXPs; the results are noisy, but the Milky Way is a $1-3$ sigma outlier in both tests, implying that perhaps the SFR of the Milky Way is overestimated and/or extragalactic SFRs are systematically underestimated.

We also test for consistency between SFR diagnostics within Galactic \hii\ regions, where we can obtain exquisite detail and actual star counts. Our comparison of SFRs based on tracers of massive stars against SFRs determined from star counts reveals that large systematic errors persist in {\it absolute} SFR determinations. In particular, measurements of SFRs in both the Milky Way and external galaxies could be underestimating absolute SFRs by factors of at least 2--3. 
Two outstanding issues may explain the discrepancy between SFR$_{\rm SC}$ and SFRs measured from indirect tracers:
\begin{enumerate}
\item {\it Uncertainty in the form of the intermediate-mass IMF.} The systematic offset would be assuaged if the IMF has a slope that is steeper than Salpeter across the super-solar mass range (1.5~\Msun$<m<5$~\Msun), a tweak that is fully consistent with current data \citep{Bastian_etal10}. 
\item {\it Systematically overestimated star-formation timescales for extragalactic \hii~regions.} The population synthesis models employed in extragalactic SFR determinations often assume that the duration of star formation is longer than an O star lifetime. This assumption breaks down in the case of \hii~regions that are mid-IR bright, because such regions are systematically younger than the O-star lifetimes by factors of ${\ga}1.6$.
\end{enumerate}
We stress that these systematic issues are not mutually exclusive, and we cannot rule out additional contributions from the O star models themselves \citep{{Smith_etal02},Martins_etal05} and runaway O stars. In addition, SFR determinations in the lowest-mass \hii\ regions ($< 3,000$ M$_{\odot}$) will also be affected by stochastic under-sampling of the high-mass end of the IMF. Because we derived only lower limits on SFR$_{\rm SC}$, the actual discrepancy may be higher than reported here. 

Nevertheless, distinguishing between the two primary scenarios is crucial to our understanding of absolute SFRs. In the IMF scenario, {\it all} Galactic and extragalactic SFR determinations must be revised upwards, with far-reaching implications for applications that rely on absolute SFRs, like the Schmidt--Kennicutt relation, galaxy evolution models, and chemical enrichment by evolved low-mass stars. For example, if SFRs were larger on galaxy-wide scales, this would increase the discrepancy between SFR integrated over cosmic time and measured stellar mass densities \citep{Wilkins_etal08, Choi_Nagamine11}. Also (assuming the Galactic SFR has been roughly constant throughout the history of the Milky Way), a larger amount of fresh un-processed ISM material would be required to infall onto the Galactic disk to explain stellar abundance patterns and sustain star formation \citep[e.g.,][]{Sancisi_etal08}. 

On the other hand, if the timescale scenario is sufficient to explain the discrepancy between SFR$_{\rm SC}$ and the calibrations based on population synthesis models in a sample of IR-bright, \hii~regions, only SFR determinations in localized, very young star-forming regions will be affected, not entire galaxies. The extragalactic H$\alpha$+IR SFR calibration of K+09, for example, accounts for the ionizing luminosity of all massive stars in a galaxy, averaging across the star formation histories of many \hii\ regions, so it should not be vulnerable to this systematic. 

Our discussion reminds us that implicit assumptions have the potential to significantly impact SFR determinations, both in a relative and absolute sense. In the future, additional tests and larger sample sizes are required to test the significance of the potential deviation of the Milky Way's SFR from extragalactic calibrations. Meanwhile, additional multiwavelength observations of Galactic star-forming regions with a range of ages will have the power to discern between the two most likely scenarios for the \hii-region SFR discrepancy. Such observations would allow for a more accurate determination of the IMF and also provide additional estimates of SFR$_{\rm SC}$ for a range of star formation histories. This should pave the way to an independent, empirical calibration of absolute SFR versus tracers of ionizing luminosity. In order to place studies of galaxy evolution, ISM physics, and stellar populations on the most solid footing possible, we need to continue toward a better understanding of the systematics affecting SFR determinations, ensuring that Galactic and extragalactic calibrations achieve convergence.


\acknowledgements
We would like to thank Ed Churchwell, Dave Green, Leisa Townsley, and Eric Wilcots for their support and useful insights. We are also grateful to our referee, Neal J.~Evans II, for his improvements to this work. M.~S.~Povich is supported by an NSF Astronomy and Astrophysics Postdoctoral Fellowship under award AST-0901646. L.~Chomiuk is a Jansky Fellow of the National Radio Astronomy Observatory. She is also grateful for the support of NSF grant AST-0708002. The National Radio Astronomy Observatory is a facility of the National Science Foundation operated under cooperative agreement by Associated Universities, Inc.

\bibliography{mwref}

\begin{thebibliography}{139}
\expandafter\ifx\csname natexlab\endcsname\relax\def\natexlab#1{#1}\fi

\bibitem[{{Alexander} {et~al.}(2009){Alexander}, {Kobulnicky}, {Clemens},
  {Jameson}, {Pinnick}, \& {Pavel}}]{Alexander_etal09}
{Alexander}, M.~J., {Kobulnicky}, H.~A., {Clemens}, D.~P., {Jameson}, K.,
  {Pinnick}, A., \& {Pavel}, M. 2009, \aj, 137, 4824

\bibitem[{{Ascenso} {et~al.}(2007){Ascenso}, {Alves}, {Beletsky}, \&
  {Lago}}]{Ascenso_etal07}
{Ascenso}, J., {Alves}, J., {Beletsky}, Y., \& {Lago}, M.~T.~V.~T. 2007, \aap,
  466, 137

\bibitem[{{Baars} {et~al.}(1977){Baars}, {Genzel}, {Pauliny-Toth}, \&
  {Witzel}}]{Baars_etal77}
{Baars}, J.~W.~M., {Genzel}, R., {Pauliny-Toth}, I.~I.~K., \& {Witzel}, A.
  1977, \aap, 61, 99

\bibitem[{{Bastian} {et~al.}(2010){Bastian}, {Covey}, \&
  {Meyer}}]{Bastian_etal10}
{Bastian}, N., {Covey}, K.~R., \& {Meyer}, M.~R. 2010, \araa, 48, 339

\bibitem[{{Bennett} {et~al.}(1994){Bennett}, {Fixsen}, {Hinshaw}, {Mather},
  {Moseley}, {Wright}, {Eplee}, {Gales}, {Hewagama}, {Isaacman}, {Shafer}, \&
  {Turpie}}]{Bennett_etal94}
{Bennett}, C.~L., {Fixsen}, D.~J., {Hinshaw}, G., {Mather}, J.~C., {Moseley},
  S.~H., {Wright}, E.~L., {Eplee}, Jr., R.~E., {Gales}, J., {Hewagama}, T.,
  {Isaacman}, R.~B., {Shafer}, R.~A., \& {Turpie}, K. 1994, \apj, 434, 587

\bibitem[{{Bertelli} {et~al.}(1994){Bertelli}, {Bressan}, {Chiosi}, {Fagotto},
  \& {Nasi}}]{Bertelli_etal94}
{Bertelli}, G., {Bressan}, A., {Chiosi}, C., {Fagotto}, F., \& {Nasi}, E. 1994,
  \aaps, 106, 275

\bibitem[{{Bouwens} {et~al.}(2011){Bouwens}, {Illingworth}, {Oesch},
  {Labb{\'e}}, {Trenti}, {van Dokkum}, {Franx}, {Stiavelli}, {Carollo},
  {Magee}, \& {Gonzalez}}]{Bouwens_etal11}
{Bouwens}, R.~J., {Illingworth}, G.~D., {Oesch}, P.~A., {Labb{\'e}}, I.,
  {Trenti}, M., {van Dokkum}, P., {Franx}, M., {Stiavelli}, M., {Carollo},
  C.~M., {Magee}, D., \& {Gonzalez}, V. 2011, \apj, 737, 90

\bibitem[{{Broos} {et~al.}(2007){Broos}, {Feigelson}, {Townsley}, {Getman},
  {Wang}, {Garmire}, {Jiang}, \& {Tsuboi}}]{Broos_etal07}
{Broos}, P.~S., {Feigelson}, E.~D., {Townsley}, L.~K., {Getman}, K.~V., {Wang},
  J., {Garmire}, G.~P., {Jiang}, Z., \& {Tsuboi}, Y. 2007, \apjs, 169, 353

\bibitem[{{Calzetti} {et~al.}(2007){Calzetti}, {Kennicutt}, {Engelbracht},
  {Leitherer}, {Draine}, {Kewley}, {Moustakas}, {Sosey}, {Dale}, {Gordon},
  {Helou}, {Hollenbach}, {Armus}, {Bendo}, {Bot}, {Buckalew}, {Jarrett}, {Li},
  {Meyer}, {Murphy}, {Prescott}, {Regan}, {Rieke}, {Roussel}, {Sheth}, {Smith},
  {Thornley}, \& {Walter}}]{Calzetti_etal07}
{Calzetti}, D., {Kennicutt}, R.~C., {Engelbracht}, C.~W., {Leitherer}, C.,
  {Draine}, B.~T., {Kewley}, L., {Moustakas}, J., {Sosey}, M., {Dale}, D.~A.,
  {Gordon}, K.~D., {Helou}, G.~X., {Hollenbach}, D.~J., {Armus}, L., {Bendo},
  G., {Bot}, C., {Buckalew}, B., {Jarrett}, T., {Li}, A., {Meyer}, M.,
  {Murphy}, E.~J., {Prescott}, M., {Regan}, M.~W., {Rieke}, G.~H., {Roussel},
  H., {Sheth}, K., {Smith}, J.~D.~T., {Thornley}, M.~D., \& {Walter}, F. 2007,
  \apj, 666, 870 (C+07)

\bibitem[{{Calzetti} {et~al.}(2009){Calzetti}, {Sheth}, {Churchwell}, \&
  {Jackson}}]{Calzetti_etal09}
{Calzetti}, D., {Sheth}, K., {Churchwell}, E., \& {Jackson}, J. 2009, in The
  Evolving ISM in the Milky Way and Nearby Galaxies, ed. {K.~Sheth,
  A.~Noriega-Crespo, J.~Ingalls, \& R.~Paladini}

\bibitem[{{Carey} {et~al.}(2009){Carey}, {Noriega-Crespo}, {Mizuno}, {Shenoy},
  {Paladini}, {Kraemer}, {Price}, {Flagey}, {Ryan}, {Ingalls}, {Kuchar},
  {Pinheiro Gon{\c c}alves}, {Indebetouw}, {Billot}, {Marleau}, {Padgett},
  {Rebull}, {Bressert}, {Ali}, {Molinari}, {Martin}, {Berriman}, {Boulanger},
  {Latter}, {Miville-Deschenes}, {Shipman}, \& {Testi}}]{Carey_etal09}
{Carey}, S.~J., {Noriega-Crespo}, A., {Mizuno}, D.~R., {Shenoy}, S.,
  {Paladini}, R., {Kraemer}, K.~E., {Price}, S.~D., {Flagey}, N., {Ryan}, E.,
  {Ingalls}, J.~G., {Kuchar}, T.~A., {Pinheiro Gon{\c c}alves}, D.,
  {Indebetouw}, R., {Billot}, N., {Marleau}, F.~R., {Padgett}, D.~L., {Rebull},
  L.~M., {Bressert}, E., {Ali}, B., {Molinari}, S., {Martin}, P.~G.,
  {Berriman}, G.~B., {Boulanger}, F., {Latter}, W.~B., {Miville-Deschenes},
  M.~A., {Shipman}, R., \& {Testi}, L. 2009, \pasp, 121, 76

\bibitem[{{Celnik}(1985)}]{Celnik85}
{Celnik}, W.~E. 1985, \aap, 144, 171

\bibitem[{{Cervi{\~n}o} {et~al.}(2003){Cervi{\~n}o}, {Luridiana}, {P{\'e}rez},
  {V{\'{\i}}lchez}, \& {Valls-Gabaud}}]{Cervino_etal03}
{Cervi{\~n}o}, M., {Luridiana}, V., {P{\'e}rez}, E., {V{\'{\i}}lchez}, J.~M.,
  \& {Valls-Gabaud}, D. 2003, \aap, 407, 177

\bibitem[{{Cervi{\~n}o} \& {Valls-Gabaud}(2003)}]{Cervino_Valls-Gabaud03}
{Cervi{\~n}o}, M. \& {Valls-Gabaud}, D. 2003, \mnras, 338, 481

\bibitem[{{Chabrier}(2003)}]{Chabrier03}
{Chabrier}, G. 2003, \pasp, 115, 763

\bibitem[{{Chabrier}(2005)}]{Chabrier05}
{Chabrier}, G. 2005, in Astrophysics and Space Science Library, Vol. 327, The
  Initial Mass Function 50 Years Later, ed. {E.~Corbelli, F.~Palla, \&
  H.~Zinnecker}, 41

\bibitem[{{Chiappini} {et~al.}(1997){Chiappini}, {Matteucci}, \&
  {Gratton}}]{Chiappini_etal97}
{Chiappini}, C., {Matteucci}, F., \& {Gratton}, R. 1997, \apj, 477, 765

\bibitem[{{Choi} \& {Nagamine}(2011)}]{Choi_Nagamine11}
{Choi}, J.-H. \& {Nagamine}, K. 2011, arXiv:1101.5656

\bibitem[{{Chomiuk} \& {Wilcots}(2009)}]{Chomiuk_Wilcots09b}
{Chomiuk}, L. \& {Wilcots}, E.~M. 2009, \apj, 703, 370 (CW09)

\bibitem[{{Chu} {et~al.}(1999){Chu}, {Chen}, \& {Lai}}]{Chu_etal99}
{Chu}, Y., {Chen}, C., \& {Lai}, S. 1999, in HST May Symposium, The Largest
  Explosions Since the Big Bang: Supernovae and Gamma Ray Bursts, arXiv:9909091

\bibitem[{{Churchwell} {et~al.}(2009){Churchwell}, {Babler}, {Meade},
  {Whitney}, {Benjamin}, {Indebetouw}, {Cyganowski}, {Robitaille}, {Povich},
  {Watson}, \& {Bracker}}]{Churchwell_etal09}
{Churchwell}, E., {Babler}, B.~L., {Meade}, M.~R., {Whitney}, B.~A.,
  {Benjamin}, R., {Indebetouw}, R., {Cyganowski}, C., {Robitaille}, T.~P.,
  {Povich}, M., {Watson}, C., \& {Bracker}, S. 2009, \pasp, 121, 213

\bibitem[{{Clark} {et~al.}(2005){Clark}, {Negueruela}, {Crowther}, \&
  {Goodwin}}]{Clark_etal05}
{Clark}, J.~S., {Negueruela}, I., {Crowther}, P.~A., \& {Goodwin}, S.~P. 2005,
  \aap, 434, 949

\bibitem[{{Collison} {et~al.}(1994){Collison}, {Saikia}, {Pedlar}, {Axon}, \&
  {Unger}}]{Collison_etal94}
{Collison}, P.~M., {Saikia}, D.~J., {Pedlar}, A., {Axon}, D.~J., \& {Unger},
  S.~W. 1994, \mnras, 268, 203

\bibitem[{{Cox} {et~al.}(1986){Cox}, {Kruegel}, \& {Mezger}}]{Cox_etal86}
{Cox}, P., {Kruegel}, E., \& {Mezger}, P.~G. 1986, \aap, 155, 380

\bibitem[{{Crowther} {et~al.}(2010){Crowther}, {Schnurr}, {Hirschi}, {Yusof},
  {Parker}, {Goodwin}, \& {Kassim}}]{Crowther_etal10}
{Crowther}, P.~A., {Schnurr}, O., {Hirschi}, R., {Yusof}, N., {Parker}, R.~J.,
  {Goodwin}, S.~P., \& {Kassim}, H.~A. 2010, \mnras, 408, 731

\bibitem[{{da Silva} {et~al.}(2011){da Silva}, {Fumagalli}, \&
  {Krumholz}}]{DaSilva_etal11}
{da Silva}, R.~L., {Fumagalli}, M., \& {Krumholz}, M. 2011, arXiv 1106.3072

\bibitem[{{Dale} \& {Helou}(2002)}]{Dale_Helou02}
{Dale}, D.~A. \& {Helou}, G. 2002, \apj, 576, 159

\bibitem[{{Davidson} \& {Fesen}(1985)}]{Davidson_Fesen85}
{Davidson}, K. \& {Fesen}, R.~A. 1985, \araa, 23, 119

\bibitem[{{Davies} {et~al.}(2007){Davies}, {Figer}, {Kudritzki}, {MacKenty},
  {Najarro}, \& {Herrero}}]{Davies_etal07}
{Davies}, B., {Figer}, D.~F., {Kudritzki}, R.-P., {MacKenty}, J., {Najarro},
  F., \& {Herrero}, A. 2007, \apj, 671, 781

\bibitem[{{Davies} {et~al.}(2011){Davies}, {Hoare}, {Lumsden}, {Hosokawa},
  {Oudmaijer}, {Urquhart}, {Mottram}, \& {Stead}}]{Davies_etal11}
{Davies}, B., {Hoare}, M.~G., {Lumsden}, S.~L., {Hosokawa}, T., {Oudmaijer},
  R.~D., {Urquhart}, J.~S., {Mottram}, J.~C., \& {Stead}, J. 2011, \mnras, 1015

\bibitem[{{de Rossi} {et~al.}(2009){de Rossi}, {Tissera}, {De Lucia}, \&
  {Kauffmann}}]{deRossi_etal09}
{de Rossi}, M.~E., {Tissera}, P.~B., {De Lucia}, G., \& {Kauffmann}, G. 2009,
  \mnras, 395, 210

\bibitem[{{de Vaucouleurs} \& {Freeman}(1972)}]{deVaucouleurs_Freeman72}
{de Vaucouleurs}, G. \& {Freeman}, K.~C. 1972, Vistas in Astronomy, 14, 163

\bibitem[{{Diehl} {et~al.}(2006){Diehl}, {Halloin}, {Kretschmer}, {Lichti},
  {Sch{\"o}nfelder}, {Strong}, {von Kienlin}, {Wang}, {Jean}, {Kn{\"o}dlseder},
  {Roques}, {Weidenspointner}, {Schanne}, {Hartmann}, {Winkler}, \&
  {Wunderer}}]{Diehl_etal06}
{Diehl}, R., {Halloin}, H., {Kretschmer}, K., {Lichti}, G.~G.,
  {Sch{\"o}nfelder}, V., {Strong}, A.~W., {von Kienlin}, A., {Wang}, W.,
  {Jean}, P., {Kn{\"o}dlseder}, J., {Roques}, J., {Weidenspointner}, G.,
  {Schanne}, S., {Hartmann}, D.~H., {Winkler}, C., \& {Wunderer}, C. 2006,
  \nat, 439, 45

\bibitem[{{Feigelson} {et~al.}(2002){Feigelson}, {Broos}, {Gaffney}, {Garmire},
  {Hillenbrand}, {Pravdo}, {Townsley}, \& {Tsuboi}}]{Feigelson_etal02}
{Feigelson}, E.~D., {Broos}, P., {Gaffney}, III, J.~A., {Garmire}, G.,
  {Hillenbrand}, L.~A., {Pravdo}, S.~H., {Townsley}, L., \& {Tsuboi}, Y. 2002,
  \apj, 574, 258

\bibitem[{{Feigelson} {et~al.}(2011){Feigelson}, {Getman}, {Townsley}, {Broos},
  {Povich}, {Garmire}, {King}, {Montmerle}, {Preibisch}, {Smith}, {Stassun},
  {Wang}, {Wolk}, \& {Zinnecker}}]{Feigelson_etal11}
{Feigelson}, E.~D., {Getman}, K.~V., {Townsley}, L.~K., {Broos}, P.~S.,
  {Povich}, M.~S., {Garmire}, G.~P., {King}, R.~R., {Montmerle}, T.,
  {Preibisch}, T., {Smith}, N., {Stassun}, K.~G., {Wang}, J., {Wolk}, S., \&
  {Zinnecker}, H. 2011, \apjs, 194, 9

\bibitem[{{Feigelson} \& {Townsley}(2008)}]{Feigelson_Townsley08}
{Feigelson}, E.~D. \& {Townsley}, L.~K. 2008, \apj, 673, 354

\bibitem[{{Figer} {et~al.}(2006){Figer}, {MacKenty}, {Robberto}, {Smith},
  {Najarro}, {Kudritzki}, \& {Herrero}}]{Figer_etal06}
{Figer}, D.~F., {MacKenty}, J.~W., {Robberto}, M., {Smith}, K., {Najarro}, F.,
  {Kudritzki}, R.~P., \& {Herrero}, A. 2006, \apj, 643, 1166

\bibitem[{{Fioc} \& {Rocca-Volmerange}(1997)}]{Fioc_Rocca97}
{Fioc}, M. \& {Rocca-Volmerange}, B. 1997, \aap, 326, 950

\bibitem[{{Fullmer} \& {Lonsdale}(1989)}]{Fullmer_Lonsdale89}
{Fullmer}, L. \& {Lonsdale}, C.~J. 1989, JPL D-1932, Version 2, part no 3

\bibitem[{{Gagn{\'e}} {et~al.}(2011){Gagn{\'e}}, {Fehon}, {Savoy}, {Cohen},
  {Townsley}, {Broos}, {Povich}, {Corcoran}, {Walborn}, {Remage Evans},
  {Moffat}, {Naz{\'e}}, \& {Oskinova}}]{Gagne_etal11}
{Gagn{\'e}}, M., {Fehon}, G., {Savoy}, M.~R., {Cohen}, D.~H., {Townsley},
  L.~K., {Broos}, P.~S., {Povich}, M.~S., {Corcoran}, M.~F., {Walborn}, N.~R.,
  {Remage Evans}, N., {Moffat}, A.~F.~J., {Naz{\'e}}, Y., \& {Oskinova}, L.~M.
  2011, \apjs, 194, 5

\bibitem[{{Getman} {et~al.}(2005){Getman}, {Feigelson}, {Grosso},
  {McCaughrean}, {Micela}, {Broos}, {Garmire}, \& {Townsley}}]{Getman_etal05}
{Getman}, K.~V., {Feigelson}, E.~D., {Grosso}, N., {McCaughrean}, M.~J.,
  {Micela}, G., {Broos}, P., {Garmire}, G., \& {Townsley}, L. 2005, \apjs, 160,
  353

\bibitem[{{Getman} {et~al.}(2006){Getman}, {Feigelson}, {Townsley}, {Broos},
  {Garmire}, \& {Tsujimoto}}]{Getman_etal06}
{Getman}, K.~V., {Feigelson}, E.~D., {Townsley}, L., {Broos}, P., {Garmire},
  G., \& {Tsujimoto}, M. 2006, \apjs, 163, 306

\bibitem[{{Gies}(1987)}]{Gies87}
{Gies}, D.~R. 1987, \apjs, 64, 545

\bibitem[{{Giles}(1977)}]{Giles77}
{Giles}, K. 1977, \mnras, 180, 57P

\bibitem[{{Gilfanov} {et~al.}(2004){Gilfanov}, {Grimm}, \&
  {Sunyaev}}]{Gilfanov_etal04}
{Gilfanov}, M., {Grimm}, H., \& {Sunyaev}, R. 2004, \mnras, 347, L57

\bibitem[{{Green}(2004)}]{Green04}
{Green}, D.~A. 2004, Bulletin of the Astronomical Society of India, 32, 335

\bibitem[{{Green}(2005)}]{Green05}
---. 2005, Memorie della Societa Astronomica Italiana, 76, 534

\bibitem[{{Grimm} {et~al.}(2002){Grimm}, {Gilfanov}, \&
  {Sunyaev}}]{Grimm_etal02}
{Grimm}, H., {Gilfanov}, M., \& {Sunyaev}, R. 2002, \aap, 391, 923

\bibitem[{{Grimm} {et~al.}(2003){Grimm}, {Gilfanov}, \&
  {Sunyaev}}]{Grimm_etal03}
{Grimm}, H.-J., {Gilfanov}, M., \& {Sunyaev}, R. 2003, \mnras, 339, 793

\bibitem[{{G\"{u}sten} \& {Mezger}(1982)}]{Guesten_Mezger82}
{G\"{u}sten}, R. \& {Mezger}, P.~G. 1982, Vistas in Astronomy, 26, 159

\bibitem[{{Gvaramadze} {et~al.}(2011){Gvaramadze}, {Kniazev}, {Kroupa}, \&
  {Oh}}]{Gvaramadze_etal11}
{Gvaramadze}, V.~V., {Kniazev}, A.~Y., {Kroupa}, P., \& {Oh}, S. 2011, A\&A,
  accepted, arXiv 1109.2116

\bibitem[{{Hammer} {et~al.}(2007){Hammer}, {Puech}, {Chemin}, {Flores}, \&
  {Lehnert}}]{Hammer_etal07}
{Hammer}, F., {Puech}, M., {Chemin}, L., {Flores}, H., \& {Lehnert}, M.~D.
  2007, \apj, 662, 322

\bibitem[{{Hanson} \& {Popescu}(2008)}]{Hanson_Popescu08}
{Hanson}, M.~M. \& {Popescu}, B. 2008, in IAU Symposium, Vol. 250, Massive
  Stars as Cosmic Engines, ed. {F.~Bresolin, P.~A.~Crowther, \& J.~Puls}, 307

\bibitem[{{Heiderman} {et~al.}(2010){Heiderman}, {Evans}, {Allen}, {Huard}, \&
  {Heyer}}]{Heiderman_etal10}
{Heiderman}, A., {Evans}, II, N.~J., {Allen}, L.~E., {Huard}, T., \& {Heyer},
  M. 2010, \apj, 723, 1019

\bibitem[{{Hillenbrand} \& {Hartmann}(1998)}]{Hillenbrand_Hartmann98}
{Hillenbrand}, L.~A. \& {Hartmann}, L.~W. 1998, \apj, 492, 540

\bibitem[{{Hunter} {et~al.}(1986){Hunter}, {Gillett}, {Gallagher}, {Rice}, \&
  {Low}}]{Hunter_etal86}
{Hunter}, D.~A., {Gillett}, F.~C., {Gallagher}, III, J.~S., {Rice}, W.~L., \&
  {Low}, F.~J. 1986, \apj, 303, 171

\bibitem[{{Jim{\'e}nez-Bail{\'o}n} {et~al.}(2005){Jim{\'e}nez-Bail{\'o}n},
  {Santos-Lle{\'o}}, {Dahlem}, {Ehle}, {Mas-Hesse}, {Guainazzi}, {Heckman}, \&
  {Weaver}}]{Jimenez-Bailon_etal05}
{Jim{\'e}nez-Bail{\'o}n}, E., {Santos-Lle{\'o}}, M., {Dahlem}, M., {Ehle}, M.,
  {Mas-Hesse}, J.~M., {Guainazzi}, M., {Heckman}, T.~M., \& {Weaver}, K.~A.
  2005, \aap, 442, 861

\bibitem[{{Kauffmann} \& {Haehnelt}(2000)}]{Kauffmann_Haehnelt00}
{Kauffmann}, G. \& {Haehnelt}, M. 2000, \mnras, 311, 576

\bibitem[{{Kennicutt} {et~al.}(2009){Kennicutt}, {Hao}, {Calzetti},
  {Moustakas}, {Dale}, {Bendo}, {Engelbracht}, {Johnson}, \&
  {Lee}}]{Kennicutt_etal09}
{Kennicutt}, R.~C., {Hao}, C., {Calzetti}, D., {Moustakas}, J., {Dale}, D.~A.,
  {Bendo}, G., {Engelbracht}, C.~W., {Johnson}, B.~D., \& {Lee}, J.~C. 2009,
  \apj, 703, 1672 (K+09)

\bibitem[{{Kennicutt}(1998{\natexlab{a}})}]{Kennicutt98a}
{Kennicutt}, Jr., R.~C. 1998{\natexlab{a}}, \araa, 36, 189

\bibitem[{{Kennicutt}(1998{\natexlab{b}})}]{Kennicutt98b}
---. 1998{\natexlab{b}}, \apj, 498, 541

\bibitem[{{Kennicutt} {et~al.}(2003){Kennicutt}, {Armus}, {Bendo}, {Calzetti},
  {Dale}, {Draine}, {Engelbracht}, {Gordon}, {Grauer}, {Helou}, {Hollenbach},
  {Jarrett}, {Kewley}, {Leitherer}, {Li}, {Malhotra}, {Regan}, {Rieke},
  {Rieke}, {Roussel}, {Smith}, {Thornley}, \& {Walter}}]{SINGS}
{Kennicutt}, Jr., R.~C., {Armus}, L., {Bendo}, G., {Calzetti}, D., {Dale},
  D.~A., {Draine}, B.~T., {Engelbracht}, C.~W., {Gordon}, K.~D., {Grauer},
  A.~D., {Helou}, G., {Hollenbach}, D.~J., {Jarrett}, T.~H., {Kewley}, L.~J.,
  {Leitherer}, C., {Li}, A., {Malhotra}, S., {Regan}, M.~W., {Rieke}, G.~H.,
  {Rieke}, M.~J., {Roussel}, H., {Smith}, J., {Thornley}, M.~D., \& {Walter},
  F. 2003, \pasp, 115, 928

\bibitem[{{Kennicutt} {et~al.}(2008){Kennicutt}, {Lee}, {Funes}, {Sakai}, \&
  {Akiyama}}]{Kennicutt_etal08}
{Kennicutt}, Jr., R.~C., {Lee}, J.~C., {Funes}, J.~G., S.~J., {Sakai}, S., \&
  {Akiyama}, S. 2008, \apjs, 178, 247

\bibitem[{{Kennicutt} {et~al.}(1994){Kennicutt}, {Tamblyn}, \&
  {Congdon}}]{Kennicutt_etal94}
{Kennicutt}, Jr., R.~C., {Tamblyn}, P., \& {Congdon}, C.~E. 1994, \apj, 435, 22

\bibitem[{{Koenig} \& {Allen}(2011)}]{Koenig_Allen11}
{Koenig}, X.~P. \& {Allen}, L.~E. 2011, \apj, 726, 18

\bibitem[{{Kroupa}(2001)}]{Kroupa01}
{Kroupa}, P. 2001, \mnras, 322, 231

\bibitem[{{Kroupa}(2004)}]{Kroupa04}
---. 2004, New Astron.~Rev, 48, 47

\bibitem[{{Kroupa} \& {Weidner}(2003)}]{Kroupa_Weidner03}
{Kroupa}, P. \& {Weidner}, C. 2003, \apj, 598, 1076

\bibitem[{{Larson}(1974)}]{Larson74}
{Larson}, R.~B. 1974, \mnras, 169, 229

\bibitem[{{Lee} {et~al.}(2009){Lee}, {Gil de Paz}, {Tremonti}, {Kennicutt},
  {Salim}, {Bothwell}, {Calzetti}, {Dalcanton}, {Dale}, {Engelbracht}, {Funes},
  {Johnson}, {Sakai}, {Skillman}, {van Zee}, {Walter}, \& {Weisz}}]{Lee_etal09}
{Lee}, J.~C., {Gil de Paz}, A., {Tremonti}, C., {Kennicutt}, Jr., R.~C.,
  {Salim}, S., {Bothwell}, M., {Calzetti}, D., {Dalcanton}, J., {Dale}, D.,
  {Engelbracht}, C., {Funes}, S.~J.~J.~G., {Johnson}, B., {Sakai}, S.,
  {Skillman}, E., {van Zee}, L., {Walter}, F., \& {Weisz}, D. 2009, \apj, 706,
  599

\bibitem[{{Lehmer} {et~al.}(2010){Lehmer}, {Alexander}, {Bauer}, {Brandt},
  {Goulding}, {Jenkins}, {Ptak}, \& {Roberts}}]{Lehmer_etal10}
{Lehmer}, B.~D., {Alexander}, D.~M., {Bauer}, F.~E., {Brandt}, W.~N.,
  {Goulding}, A.~D., {Jenkins}, L.~P., {Ptak}, A., \& {Roberts}, T.~P. 2010,
  \apj, 724, 559

\bibitem[{{Leitherer} {et~al.}(1999){Leitherer}, {Schaerer}, {Goldader},
  {Gonz{\'a}lez Delgado}, {Robert}, {Kune}, {de Mello}, {Devost}, \&
  {Heckman}}]{Leitherer_etal99}
{Leitherer}, C., {Schaerer}, D., {Goldader}, J.~D., {Gonz{\'a}lez Delgado},
  R.~M., {Robert}, C., {Kune}, D.~F., {de Mello}, D.~F., {Devost}, D., \&
  {Heckman}, T.~M. 1999, \apjs, 123, 3

\bibitem[{{Leonard} \& {Duncan}(1990)}]{Leonard_Duncan90}
{Leonard}, P.~J.~T. \& {Duncan}, M.~J. 1990, \aj, 99, 608

\bibitem[{{Limongi} \& {Chieffi}(2006)}]{Limongi_Chieffi06}
{Limongi}, M. \& {Chieffi}, A. 2006, \apj, 647, 483

\bibitem[{{Lozinskaya} \& {Moiseev}(2007)}]{Lozinskaya_Moiseev07}
{Lozinskaya}, T.~A. \& {Moiseev}, A.~V. 2007, \mnras, 381, L26

\bibitem[{{Madau} {et~al.}(1996){Madau}, {Ferguson}, {Dickinson}, {Giavalisco},
  {Steidel}, \& {Fruchter}}]{Madau_etal96}
{Madau}, P., {Ferguson}, H.~C., {Dickinson}, M.~E., {Giavalisco}, M.,
  {Steidel}, C.~C., \& {Fruchter}, A. 1996, \mnras, 283, 1388

\bibitem[{{Ma{\'{\i}}z Apell{\'a}niz}(2008)}]{MA08}
{Ma{\'{\i}}z Apell{\'a}niz}, J. 2008, \apj, 677, 1278

\bibitem[{{Ma{\'{\i}}z Apell{\'a}niz} {et~al.}(2007){Ma{\'{\i}}z
  Apell{\'a}niz}, {Walborn}, {Morrell}, {Niemela}, \&
  {Nelan}}]{Maiz_Apellaniz_etal07}
{Ma{\'{\i}}z Apell{\'a}niz}, J., {Walborn}, N.~R., {Morrell}, N.~I., {Niemela},
  V.~S., \& {Nelan}, E.~P. 2007, \apj, 660, 1480

\bibitem[{{Martins} {et~al.}(2005){Martins}, {Schaerer}, \&
  {Hillier}}]{Martins_etal05}
{Martins}, F., {Schaerer}, D., \& {Hillier}, D.~J. 2005, \aap, 436, 1049

\bibitem[{{McKee} \& {Williams}(1997)}]{McKee_Williams97}
{McKee}, C.~F. \& {Williams}, J.~P. 1997, \apj, 476, 144

\bibitem[{{Mezger}(1987)}]{Mezger87}
{Mezger}, P.~G. 1987, in Starbursts and Galaxy Evolution, ed. {T.~X.~Thuan,
  T.~Montmerle, \& J.~Tran Thanh van}, 3

\bibitem[{{Mezger} {et~al.}(1974){Mezger}, {Smith}, \&
  {Churchwell}}]{Mezger_etal74}
{Mezger}, P.~G., {Smith}, L.~F., \& {Churchwell}, E. 1974, \aap, 32, 269

\bibitem[{{Milisavljevic} \& {Fesen}(2008)}]{Milisavljevic_Fesen08}
{Milisavljevic}, D. \& {Fesen}, R.~A. 2008, \apj, 677, 306

\bibitem[{{Miller} \& {Scalo}(1979)}]{Miller_Scalo79}
{Miller}, G.~E. \& {Scalo}, J.~M. 1979, \apjs, 41, 513

\bibitem[{{Mineo} {et~al.}(2011){Mineo}, {Gilfanov}, \&
  {Sunyaev}}]{Mineo_etal11}
{Mineo}, S., {Gilfanov}, M., \& {Sunyaev}, R. 2011, arXiv 1105.4610

\bibitem[{{Misiriotis} {et~al.}(2004){Misiriotis}, {Papadakis}, {Kylafis}, \&
  {Papamastorakis}}]{Misiriotis_etal04}
{Misiriotis}, A., {Papadakis}, I.~E., {Kylafis}, N.~D., \& {Papamastorakis}, J.
  2004, \aap, 417, 39

\bibitem[{{Misiriotis} {et~al.}(2006){Misiriotis}, {Xilouris},
  {Papamastorakis}, {Boumis}, \& {Goudis}}]{Misiriotis_etal06}
{Misiriotis}, A., {Xilouris}, E.~M., {Papamastorakis}, J., {Boumis}, P., \&
  {Goudis}, C.~D. 2006, \aap, 459, 113

\bibitem[{{Moffat} {et~al.}(1998){Moffat}, {Marchenko}, {Seggewiss}, {van der
  Hucht}, {Schrijver}, {Stenholm}, {Lundstrom}, {Setia Gunawan}, {Sutantyo},
  {van den Heuvel}, {de Cuyper}, \& {Gomez}}]{Moffat_etal98}
{Moffat}, A.~F.~J., {Marchenko}, S.~V., {Seggewiss}, W., {van der Hucht},
  K.~A., {Schrijver}, H., {Stenholm}, B., {Lundstrom}, I., {Setia Gunawan},
  D.~Y.~A., {Sutantyo}, W., {van den Heuvel}, E.~P.~J., {de Cuyper}, J.-P., \&
  {Gomez}, A.~E. 1998, \aap, 331, 949

\bibitem[{{Moshir} {et~al.}(1992){Moshir}, {Kopman}, \&
  {Conrow}}]{Moshir_etal92}
{Moshir}, M., {Kopman}, G., \& {Conrow}, T.~A.~O. 1992, {IRAS Faint Source
  Survey, Explanatory supplement version 2} (Pasadena: Infrared Processing and
  Analysis Center, California Institute of Technology)

\bibitem[{{Moustakas} \& {Kennicutt}(2006)}]{Moustakas_Kennicutt06}
{Moustakas}, J. \& {Kennicutt}, Jr., R.~C. 2006, \apjs, 164, 81

\bibitem[{{Muench} {et~al.}(2002){Muench}, {Lada}, {Lada}, \&
  {Alves}}]{Muench_etal02}
{Muench}, A.~A., {Lada}, E.~A., {Lada}, C.~J., \& {Alves}, J. 2002, \apj, 573,
  366

\bibitem[{{Murray} \& {Rahman}(2010)}]{Murray_Rahman10}
{Murray}, N. \& {Rahman}, M. 2010, \apj, 709, 424

\bibitem[{{Neff} \& {Ulvestad}(2000)}]{Neff_Ulvestad00}
{Neff}, S.~G. \& {Ulvestad}, J.~S. 2000, \aj, 120, 670

\bibitem[{{Neugebauer} {et~al.}(1984){Neugebauer}, {Habing}, {van Duinen},
  {Aumann}, {Baud}, {Beichman}, {Beintema}, {Boggess}, {Clegg}, {de Jong},
  {Emerson}, {Gautier}, {Gillett}, {Harris}, {Hauser}, {Houck}, {Jennings},
  {Low}, {Marsden}, {Miley}, {Olnon}, {Pottasch}, {Raimond}, {Rowan-Robinson},
  {Soifer}, {Walker}, {Wesselius}, \& {Young}}]{IRAS}
{Neugebauer}, G., {Habing}, H.~J., {van Duinen}, R., {Aumann}, H.~H., {Baud},
  B., {Beichman}, C.~A., {Beintema}, D.~A., {Boggess}, N., {Clegg}, P.~E., {de
  Jong}, T., {Emerson}, J.~P., {Gautier}, T.~N., {Gillett}, F.~C., {Harris},
  S., {Hauser}, M.~G., {Houck}, J.~R., {Jennings}, R.~E., {Low}, F.~J.,
  {Marsden}, P.~L., {Miley}, G., {Olnon}, F.~M., {Pottasch}, S.~R., {Raimond},
  E., {Rowan-Robinson}, M., {Soifer}, B.~T., {Walker}, R.~G., {Wesselius},
  P.~R., \& {Young}, E. 1984, \apjl, 278, L1

\bibitem[{{Oey} \& {Lamb}(2011)}]{Oey_Lamb11}
{Oey}, M.~S. \& {Lamb}, J.~B. 2011, arXiv 1109.0759

\bibitem[{{Osterbrock}(1974)}]{Osterbrock74}
{Osterbrock}, D.~E. 1974, {Astrophysics of Gaseous Nebulae} ({W. H. Freeman and
  Co.})

\bibitem[{{Pakull} \& {Gris{\'e}}(2008)}]{Pakull_Grise08}
{Pakull}, M.~W. \& {Gris{\'e}}, F. 2008, in AIP Conf. Ser., Vol. 1010, A
  Population Explosion: The Nature \& Evolution of X-ray Binaries in Diverse
  Environments, ed. {R.~M.~Bandyopadhyay, S.~Wachter, D.~Gelino, \&
  C.~R.~Gelino}, 303

\bibitem[{{Paladini} {et~al.}(2003){Paladini}, {Burigana}, {Davies}, {Maino},
  {Bersanelli}, {Cappellini}, {Platania}, \& {Smoot}}]{Paladini_etal03}
{Paladini}, R., {Burigana}, C., {Davies}, R.~D., {Maino}, D., {Bersanelli}, M.,
  {Cappellini}, B., {Platania}, P., \& {Smoot}, G. 2003, \aap, 397, 213

\bibitem[{{Panagia}(1973)}]{Panagia73}
{Panagia}, N. 1973, \aj, 78, 929

\bibitem[{{Pannuti} {et~al.}(2002){Pannuti}, {Duric}, {Lacey}, {Ferguson},
  {Magnor}, \& {Mendelowitz}}]{Pannuti_etal02}
{Pannuti}, T.~G., {Duric}, N., {Lacey}, C.~K., {Ferguson}, A.~M.~N., {Magnor},
  M.~A., \& {Mendelowitz}, C. 2002, \apj, 565, 966

\bibitem[{{Parra} {et~al.}(2007){Parra}, {Conway}, {Diamond}, {Thrall},
  {Lonsdale}, {Lonsdale}, \& {Smith}}]{Parra_etal07}
{Parra}, R., {Conway}, J.~E., {Diamond}, P.~J., {Thrall}, H., {Lonsdale},
  C.~J., {Lonsdale}, C.~J., \& {Smith}, H.~E. 2007, \apj, 659, 314

\bibitem[{{P{\'e}rez-Torres} {et~al.}(2009){P{\'e}rez-Torres},
  {Romero-Ca{\~n}izales}, {Alberdi}, \& {Polatidis}}]{Perez-Torres_etal09}
{P{\'e}rez-Torres}, M.~A., {Romero-Ca{\~n}izales}, C., {Alberdi}, A., \&
  {Polatidis}, A. 2009, \aap, 507, L17

\bibitem[{{Persic} \& {Rephaeli}(2007)}]{Persic_Rephaeli07}
{Persic}, M. \& {Rephaeli}, Y. 2007, \aap, 463, 481

\bibitem[{{Povich} {et~al.}(2008){Povich}, {Benjamin}, {Whitney}, {Babler},
  {Indebetouw}, {Meade}, \& {Churchwell}}]{Povich_etal08}
{Povich}, M.~S., {Benjamin}, R.~A., {Whitney}, B.~A., {Babler}, B.~L.,
  {Indebetouw}, R., {Meade}, M.~R., \& {Churchwell}, E. 2008, \apj, 689, 242

\bibitem[{{Povich} {et~al.}(2009){Povich}, {Churchwell}, {Bieging}, {Kang},
  {Whitney}, {Brogan}, {Kulesa}, {Cohen}, {Babler}, {Indebetouw}, {Meade}, \&
  {Robitaille}}]{Povich_etal09}
{Povich}, M.~S., {Churchwell}, E., {Bieging}, J.~H., {Kang}, M., {Whitney},
  B.~A., {Brogan}, C.~L., {Kulesa}, C.~A., {Cohen}, M., {Babler}, B.~L.,
  {Indebetouw}, R., {Meade}, M.~R., \& {Robitaille}, T.~P. 2009, \apj, 696,
  1278

\bibitem[{{Povich} {et~al.}(2011){Povich}, {Smith}, {Majewski}, {Getman},
  {Townsley}, {Babler}, {Broos}, {Indebetouw}, {Meade}, {Robitaille},
  {Stassun}, {Whitney}, {Yonekura}, \& {Fukui}}]{Povich_etal11}
{Povich}, M.~S., {Smith}, N., {Majewski}, S.~R., {Getman}, K.~V., {Townsley},
  L.~K., {Babler}, B.~L., {Broos}, P.~S., {Indebetouw}, R., {Meade}, M.~R.,
  {Robitaille}, T.~P., {Stassun}, K.~G., {Whitney}, B.~A., {Yonekura}, Y., \&
  {Fukui}, Y. 2011, \apjs, 194, 14

\bibitem[{{Povich} {et~al.}(2007){Povich}, {Stone}, {Churchwell}, {Zweibel},
  {Wolfire}, {Babler}, {Indebetouw}, {Meade}, \& {Whitney}}]{Povich_etal07}
{Povich}, M.~S., {Stone}, J.~M., {Churchwell}, E., {Zweibel}, E.~G., {Wolfire},
  M.~G., {Babler}, B.~L., {Indebetouw}, R., {Meade}, M.~R., \& {Whitney}, B.~A.
  2007, \apj, 660, 346

\bibitem[{{Povich} \& {Whitney}(2010)}]{Povich_Whitney10}
{Povich}, M.~S. \& {Whitney}, B.~A. 2010, \apjl, 714, L285

\bibitem[{{Preibisch} \& {Feigelson}(2005)}]{Preibisch_Feigelson05}
{Preibisch}, T. \& {Feigelson}, E.~D. 2005, \apjs, 160, 390

\bibitem[{{Price} {et~al.}(2001){Price}, {Egan}, {Carey}, {Mizuno}, \&
  {Kuchar}}]{Price_etal01}
{Price}, S.~D., {Egan}, M.~P., {Carey}, S.~J., {Mizuno}, D.~R., \& {Kuchar},
  T.~A. 2001, \aj, 121, 2819

\bibitem[{{Ranalli} {et~al.}(2003){Ranalli}, {Comastri}, \&
  {Setti}}]{Ranalli_etal03}
{Ranalli}, P., {Comastri}, A., \& {Setti}, G. 2003, \aap, 399, 39

\bibitem[{{Reed}(2005)}]{Reed05}
{Reed}, B.~C. 2005, \aj, 130, 1652

\bibitem[{{Reed} {et~al.}(1995){Reed}, {Hester}, {Fabian}, \&
  {Winkler}}]{Reed_etal95}
{Reed}, J.~E., {Hester}, J.~J., {Fabian}, A.~C., \& {Winkler}, P.~F. 1995,
  \apj, 440, 706

\bibitem[{{Rice} {et~al.}(1988){Rice}, {Lonsdale}, {Soifer}, {Neugebauer},
  {Kopan}, {Lloyd}, {de Jong}, \& {Habing}}]{Rice_etal88}
{Rice}, W., {Lonsdale}, C.~J., {Soifer}, B.~T., {Neugebauer}, G., {Kopan},
  E.~L., {Lloyd}, L.~A., {de Jong}, T., \& {Habing}, H.~J. 1988, \apjs, 68, 91

\bibitem[{{Robitaille} {et~al.}(2008){Robitaille}, {Meade}, {Babler},
  {Whitney}, {Johnston}, {Indebetouw}, {Cohen}, {Povich}, {Sewilo}, {Benjamin},
  \& {Churchwell}}]{Robitaille_etal08}
{Robitaille}, T.~P., {Meade}, M.~R., {Babler}, B.~L., {Whitney}, B.~A.,
  {Johnston}, K.~G., {Indebetouw}, R., {Cohen}, M., {Povich}, M.~S., {Sewilo},
  M., {Benjamin}, R.~A., \& {Churchwell}, E. 2008, \aj, 136, 2413

\bibitem[{{Robitaille} \& {Whitney}(2010)}]{Robitaille_Whitney10a}
{Robitaille}, T.~P. \& {Whitney}, B.~A. 2010, \apjl, 710, L11

\bibitem[{{Russeil} {et~al.}(2010){Russeil}, {Zavagno}, {Motte}, {Schneider},
  {Bontemps}, \& {Walsh}}]{Russeil_etal10}
{Russeil}, D., {Zavagno}, A., {Motte}, F., {Schneider}, N., {Bontemps}, S., \&
  {Walsh}, A.~J. 2010, \aap, 515, A55+

\bibitem[{{Saikia} {et~al.}(1990){Saikia}, {Unger}, {Pedlar}, {Yates}, {Axon},
  {Wolstencroft}, {Taylor}, \& {Gyldenkerne}}]{Saikia_etal90}
{Saikia}, D.~J., {Unger}, S.~W., {Pedlar}, A., {Yates}, G.~J., {Axon}, D.~J.,
  {Wolstencroft}, R.~D., {Taylor}, K., \& {Gyldenkerne}, K. 1990, \mnras, 245,
  397

\bibitem[{{Salpeter}(1955)}]{Salpeter55}
{Salpeter}, E.~E. 1955, \apj, 121, 161

\bibitem[{{Sana} {et~al.}(2008){Sana}, {Gosset}, {Naz{\'e}}, {Rauw}, \&
  {Linder}}]{Sana_etal08}
{Sana}, H., {Gosset}, E., {Naz{\'e}}, Y., {Rauw}, G., \& {Linder}, N. 2008,
  \mnras, 386, 447

\bibitem[{{Sancisi} {et~al.}(2008){Sancisi}, {Fraternali}, {Oosterloo}, \& {van
  der Hulst}}]{Sancisi_etal08}
{Sancisi}, R., {Fraternali}, F., {Oosterloo}, T., \& {van der Hulst}, T. 2008,
  \aapr, 15, 189

\bibitem[{{Sanders} {et~al.}(2003){Sanders}, {Mazzarella}, {Kim}, {Surace}, \&
  {Soifer}}]{Sanders_etal03}
{Sanders}, D.~B., {Mazzarella}, J.~M., {Kim}, D.-C., {Surace}, J.~A., \&
  {Soifer}, B.~T. 2003, \aj, 126, 1607

\bibitem[{{Scalo}(1998)}]{Scalo98}
{Scalo}, J. 1998, in Astronomical Society of the Pacific Conference Series,
  Vol. 142, The Stellar Initial Mass Function (38th Herstmonceux Conference),
  ed. {G.~Gilmore \& D.~Howell}, 201

\bibitem[{{Siess} {et~al.}(2000){Siess}, {Dufour}, \&
  {Forestini}}]{Siess_etal00}
{Siess}, L., {Dufour}, E., \& {Forestini}, M. 2000, \aap, 358, 593

\bibitem[{{Smith} {et~al.}(1978){Smith}, {Biermann}, \&
  {Mezger}}]{Smith_etal78}
{Smith}, L.~F., {Biermann}, P., \& {Mezger}, P.~G. 1978, \aap, 66, 65

\bibitem[{{Smith} {et~al.}(2002){Smith}, {Norris}, \&
  {Crowther}}]{Smith_etal02}
{Smith}, L.~J., {Norris}, R.~P.~F., \& {Crowther}, P.~A. 2002, \mnras, 337,
  1309

\bibitem[{{Smith} \& {Brooks}(2007)}]{Smith_Brooks07}
{Smith}, N. \& {Brooks}, K.~J. 2007, \mnras, 379, 1279

\bibitem[{{Tsujimoto} {et~al.}(2007){Tsujimoto}, {Feigelson}, {Townsley},
  {Broos}, {Getman}, {Wang}, {Garmire}, {Baba}, {Nagayama}, {Tamura}, \&
  {Churchwell}}]{Tsujimoto_etal07}
{Tsujimoto}, M., {Feigelson}, E.~D., {Townsley}, L.~K., {Broos}, P.~S.,
  {Getman}, K.~V., {Wang}, J., {Garmire}, G.~P., {Baba}, D., {Nagayama}, T.,
  {Tamura}, M., \& {Churchwell}, E.~B. 2007, \apj, 665, 719

\bibitem[{{Ulvestad}(2009)}]{Ulvestad09}
{Ulvestad}, J.~S. 2009, \aj, 138, 1529

\bibitem[{{Vacca} {et~al.}(1996){Vacca}, {Garmany}, \& {Shull}}]{Vacca_etal96}
{Vacca}, W.~D., {Garmany}, C.~D., \& {Shull}, J.~M. 1996, \apj, 460, 914

\bibitem[{{V{\'a}zquez} \& {Leitherer}(2005)}]{Vazquez_Leitherer05}
{V{\'a}zquez}, G.~A. \& {Leitherer}, C. 2005, \apj, 621, 695

\bibitem[{{Wang} {et~al.}(2008){Wang}, {Townsley}, {Feigelson}, {Broos},
  {Getman}, {Rom{\'a}n-Z{\'u}{\~n}iga}, \& {Lada}}]{Wang_etal08}
{Wang}, J., {Townsley}, L.~K., {Feigelson}, E.~D., {Broos}, P.~S., {Getman},
  K.~V., {Rom{\'a}n-Z{\'u}{\~n}iga}, C.~G., \& {Lada}, E. 2008, \apj, 675, 464

\bibitem[{{Wang} {et~al.}(2007){Wang}, {Townsley}, {Feigelson}, {Getman},
  {Broos}, {Garmire}, \& {Tsujimoto}}]{Wang_etal07}
{Wang}, J., {Townsley}, L.~K., {Feigelson}, E.~D., {Getman}, K.~V., {Broos},
  P.~S., {Garmire}, G.~P., \& {Tsujimoto}, M. 2007, \apjs, 168, 100

\bibitem[{{Weidner} \& {Vink}(2010)}]{Weidner_Vink10}
{Weidner}, C. \& {Vink}, J.~S. 2010, \aap, 524, A98

\bibitem[{{Wilkins} {et~al.}(2008){Wilkins}, {Trentham}, \&
  {Hopkins}}]{Wilkins_etal08}
{Wilkins}, S.~M., {Trentham}, N., \& {Hopkins}, A.~M. 2008, \mnras, 385, 687

\bibitem[{{Wolk} {et~al.}(2006){Wolk}, {Spitzbart}, {Bourke}, \&
  {Alves}}]{Wolk_etal06}
{Wolk}, S.~J., {Spitzbart}, B.~D., {Bourke}, T.~L., \& {Alves}, J. 2006, \aj,
  132, 1100

\bibitem[{{Xu} {et~al.}(2011){Xu}, {Moscadelli}, {Reid}, {Menten}, {Zhang},
  {Zheng}, \& {Brunthaler}}]{Xu_etal11}
{Xu}, Y., {Moscadelli}, L., {Reid}, M.~J., {Menten}, K.~M., {Zhang}, B.,
  {Zheng}, X.~W., \& {Brunthaler}, A. 2011, \apj, 733, 25

\bibitem[{{Yang} \& {Skillman}(1993)}]{Yang_Skillman93}
{Yang}, H. \& {Skillman}, E.~D. 1993, \aj, 106, 1448

\bibitem[{{Zinnecker} \& {Yorke}(2007)}]{Zinnecker_Yorke07}
{Zinnecker}, H. \& {Yorke}, H.~W. 2007, \araa, 45, 481

\end{thebibliography}

\appendix
\section{Details of the $L_{\rm max}$--SFR relation} 
 In an attempt to include as many galaxies as possible on the $L_{\rm max}$--SFR relation described in \S\ref{casa}, we compile SNRs at 4.85 GHz, as opposed to the 1.45 GHz data used in CW09. Free-free and sychrotron-self absorption can greatly reduce the lower frequency ($\sim$1--2 GHz) radio continuum signal coming from SNRs in very dense, violently star-forming systems. However, frequencies around $\sim$5 GHz remain unabsorbed \citep[e.g.,][]{Parra_etal07}, and are therefore ideal for searching for SNRs in starbursts. 4.85 GHz SNR samples are selected by the criteria described in CW09 and from the same references, with the addition of a few galaxies: NGC 1808 (radio data from \citealt{Saikia_etal90} and \citealt{Collison_etal94}; excluding 58.56$-$36.3 which is probably the AGN nucleus of the galaxy; \citealt{Jimenez-Bailon_etal05}); NGC 4038/4039 \citep{Neff_Ulvestad00}; and Arp 299 (\citealt{Perez-Torres_etal09} and \citealt{Ulvestad09}). In contrast with CW09, here we include the superbubble in IC 10. Although it is a severe outlier, being unusually luminous for a galaxy with IC 10's SFR, we cannot exclude it on these grounds alone as it might have an important impact on the $L_{\rm max}$--SFR relation.

To calculate the luminosity of Cas A, we use a a 4.85 GHz flux density of 806 Jy \citep{Baars_etal77}. The luminosity of Cas A is fading with time (by $\sim$1\% per year; \citealt{Baars_etal77}, and these flux densities are standardized to epoch 1980.0. We use a distance of 3.4 kpc to Cas A \citep{Reed_etal95}. The fact that the luminosity of Cas A is decreasing implies that the Milky Way's position on the $L_{\rm max}$--SFR relation will change with time, but we note that this is also likely true for the other galaxies on the relation. This fact underscores the point that the $L_{\rm max}$--SFR relation is a statistical and stochastic relation; for it to hold true, as the most luminous SNR fades, another must be born and rise to take its place.

Shortly after the analysis of CW09 was published, a new calibration for H$\alpha$+IR SFRs was released by K+09. Unlike the previous C+07 SFR calibration, the K+09 formulation is calibrated to entire galaxies, not individual star-forming regions. To calculate the extragalactic SFRs, we use the H$\alpha$+total IR (TIR) calibration from K+09, described as:  
\begin{equation} \label{sfr_k09}
\textrm{SFR}\ (\textrm{M}_{\odot}\ \textrm{yr}^{-1}) = 5.5 \times 10^{-42}\ (L_{\rm H\alpha} + 0.0024\ L_{\rm TIR} )
\end{equation} 
where L$_{H\alpha}$ and L$_{TIR}$ are both in units of erg s$^{-1}$, L$_{H\alpha}$ is corrected for foreground Galactic extinction, and L$_{TIR}$ is a linear combination of luminosity measurements in the {\it Infrared Astronomical Satellite (IRAS)} 25, 60, and 100 $\mu$m bands as suggested by \cite{Dale_Helou02}. Distances and H$\alpha$ measurements come from \cite{Kennicutt_etal08} if available, and if not, from \cite{Moustakas_Kennicutt06}. {\it IRAS} measurements come predominantly from \cite{Sanders_etal03}, but also from \cite{Fullmer_Lonsdale89}, \cite{Moshir_etal92}, \cite{Hunter_etal86}, and \cite{Rice_etal88}. Basic information on the sample galaxies and derived SFRs can be found in Table \ref{tab:gals}. 

Table \ref{tab:gals} also lists the number of SNRs (N$_{SNR}$) in the complete samples at 4.85 GHz. Many SNR surveys do not cover the entire extent of the galaxy, so we also include the field of view of the SNR survey (FoV), and the fraction of the SFR covered in this field of view (SFR Frac). The SFRs used in the $L_{\rm max}$--SFR relation are the measured SFRs times these fractions.

As in CW09, we find that the SNR LF can be described as a power law: $n(L) = A\ L^{\beta}$. Using the techniques described in CW09, we find, for the 4.85 GHz data, a best-fit power-law index of $\beta = -2.02$ and near-linear scaling of the LF with SFR as:
\begin{equation} \label{eq:a_sfr5}
A_{4.85} = \left(75^{+19}_{-15}\right) \textrm{SFR}^{1.19\pm0.11}.
\end{equation}
These measurements of the SNR LF allow us to quantify the fact that galaxies with higher SFRs are more likely to host very luminous SNRs because they have larger populations of SNRs. 

We can quantify the probability that the Galactic SFR converged upon in \S\ref{mwdiag} is consistent with the by $L_{\rm max}$--SFR relation by assuming that: (a) the $L_{\rm max}$--SFR relation is the result of stochastic sampling of a power law; and (b) the power law scaling can be described by Equation \ref{eq:a_sfr5}. For a given SFR, we then calculate how many SNRs can be expected for a power law with this scaling, and Monte-Carlo sample that number of SNRs from the power law distribution 10$^{4}$ times. This then provides 10$^{4}$ values of $L_{\rm max}$ for the given SFR, and we repeat this process for a range of SFRs from 0.005 to 20 M$_{\odot}$ yr$^{-1}$. For each SFR, we measure what fraction of these $L_{\rm max}$ values fall within a factor of 1.5 of the luminosity of Cas A. This probability as a function of SFR has a roughly log-normal form, peaking at 0.2 M$_{\odot}$ yr$^{-1}$.


If we use the 1.4 GHz data presented in CW09 (which has a smaller sample size than the 4.85 GHz data) for an analysis of the Milky Way on the $L_{\rm max}$--SFR relation, we find that the Milky Way is offset from the correlation in the same direction, although it is less aberrant (giving a 15\% chance that the SFR of the Milky Way is $1.9 \pm 0.4$ M$_{\odot}$ yr$^{-1}$). 

\begin{deluxetable}{lccccccccccc}
\tabletypesize{\footnotesize}
\tablewidth{0 pt}
\tablecaption{ \label{tab:gals}
 Galaxies Used in the $L_{\rm max}$--SFR Relation for SNRs}
\tablehead{Galaxy & R.A. (2000)\tablenotemark{a}& Dec (2000)\tablenotemark{a} & Type\tablenotemark{a} & Distance & SFR & SFR Frac & FoV  & N$_{\rm SNR}$ \\
& (hr:min:sec) & ($^{\circ}$:\arcmin:\arcsec) & & (Mpc) & (M$_{\odot}$ yr$^{-1}$) & & (arcmin) & (4.85 GHz) }
\startdata
LMC & 05:23:34.5 & $-$69:45:22 & SB(s)m & $0.05\pm0.003$ & 0.21 & 1.0 & 600 & 26 \\
SMC & 00:52:44.8 & $-$72:49:43 & SB(s)m pec & $0.06\pm0.003$ & 0.03 & 1.0 & 270 & 11 \\
IC 10 & 00:20:17.3 & +59:18:14  & IBm\tablenotemark{b} & $0.66\pm0.03$ & 0.02 & 0.91 & 4.3 & 4 \\
M 33 & 01:33:50.9 & +30:39:36 & SA(s)cd & $0.84\pm0.04$ & 0.25 & 0.98 & 40 & 35 \\
NGC 1569 & 04:30:49.0 & +64:50:53 & IBm & $1.9\pm0.4$ & 0.24 & 1.0 & 2.2 & 18 \\
NGC 300 & 00:54:53.5 & $-$37:41:04 & SA(s)d & $2.00\pm0.10$ & 0.10 & 0.65 & 9 & 8 \\
NGC 4214 & 12:15:39.2 & +36:19:37 & IAB(s)m & $2.92\pm0.29$ & 0.15 & 0.77 & 2.2 & 3 \\
NGC 2366 & 07:28:54.6 & +69:12:57 & IB(s)m & $3.19\pm0.32$ & 0.08 & 1.0 & 4.6 & 4 \\
M 82 & 09:55:52.7 & +69:40:46 & I0 & $3.53\pm0.35$ & 3.51 & 1.0 & 0.5 & 23 \\
M 81 & 09:55:33.2 & +69:03:55 & SA(s)ab & $3.63\pm0.18$ & 0.52 & 0.88 & 12.5 & 3 \\
NGC 7793 & 23:57:49.8 & $-$32:35:28 & SA(s)d & $3.91\pm0.39$ & 0.28 & 1.0 & 9 & 4 \\
NGC 253 & 00:47:33.1 & $-$25:17:18 & SAB(s)c & $3.94\pm0.39$ & 2.89 & 0.69 & 2.75 & 18 \\
NGC 4449 & 12:28:11.9 & +44:05:40 & IBm & $4.21\pm0.42$ & 0.41 & 1.0 & 5 & 5 \\
M 83 & 13:37:00.9 & $-$29:51:56 & SAB(s)c & $4.47\pm0.22$ & 2.08 & 0.98 & 9 & 14 \\
NGC 4736 & 12:50:53.0 & +41:07:14 & (R)SA(r)ab & $4.66\pm0.47$ & 0.39 & 0.81 & 2 & 10 \\
NGC 6946 & 20:34:52.3 & +60:09:14 & SAB(rs)cd & $5.9\pm1.2$ & 2.61 & 0.99 & 9  & 21 \\
NGC 4258 & 12:18:57.5 & +47:18:14 & SAB(s)bc & $7.98\pm0.40$ & 1.33 & 0.79 & 7 & 4 \\
M 51 & 13:29:55.7 & +47:13:53 & SAbc & $8.00\pm1.04$ & 2.30 & 1.0 & 9 & 30 \\
NGC 1808 & 05:07:42.3 & $-$37:30:47 & (R)SAB(s)b & $10.6\pm2.1$ & 2.52 & 0.60 & 0.3 & 10 \\
NGC 4038/39 & 12:01:53.3 & $-$18:52:37 & Pec & $25.0\pm5.0$ & 10.4 & 1.0 & 2 & 35 \\
Arp 299 & 11:28:32.2 & +58:33:44 & Pec & $50.6\pm10.1$ & 56.0 & 0.5 & 0.01 & 19 \\
Arp 220 & 15:34:57.1 & +23:30:11 & Pec & $85.2\pm17.0$ & 77.2 & 1.0 & 0.02 & 12\\
\enddata
\tablenotetext{a}{From NED.\ $^b$From \cite{deVaucouleurs_Freeman72}.}
\end{deluxetable}

\end{document}